\documentclass{ifacconf}

\usepackage{graphicx}      
\usepackage{natbib}        
\usepackage{nomencl}
\usepackage{amsmath}
\usepackage{amssymb}
\usepackage[disable]{todonotes}
\usepackage{graphicx}
\usepackage{acro}
\usepackage{tikz}
\usepackage{tikz-uml}
\usepackage[per-mode=fraction]{siunitx}

\graphicspath{{images/}}

\makeatletter
\def\endfigure{\end@float}
\def\endtable{\end@float}
\makeatother

\usepackage[caption=false]{subfig}
\usepackage{etoolbox}

\AtBeginEnvironment{figure}{\setlength{\captionwidth}{0.8\linewidth}}
\AtBeginEnvironment{table}{\setlength{\captionwidth}{0.8\linewidth}}
\newcommand{\T}[1]{\boldsymbol{\mathrm{#1}}}

\newcommand{\circledarrow}[3]{
    \draw[#1,-latex] (#2) +(-90:#3) arc(-90:180:#3);
}

\newcommand{\blockArrow}[3][]{
    \filldraw[#1] (#2) -- +(-#3/3,-#3/3) -- +(-#3/3,-#3/6) -- +(-#3,-#3/6) -- +(-#3,#3/6) -- +(-#3/3,#3/6) -- +(-#3/3,#3/3) -- cycle;
}

\newtheorem{remark}{Remark}

\definecolor{TUMBlue}{RGB}{0,101,189}
\definecolor{TUMOrange}{RGB}{227,114,34}
\definecolor{TUMGreenOriginal}{RGB}{162,173,0}
\colorlet{TUMGreen}{TUMGreenOriginal!120} 
\definecolor{TUMRose}{RGB}{227,130,143}
\definecolor{TUMGrayMedium}{RGB}{156,157,159}

\DeclareAcronym{dbc}{
    short = DBC,
    long = Dirichlet boundary condition,
}

\DeclareAcronym{nbc}{
    short = NBC,
    long = Neumann boundary condition,
}

\DeclareAcronym{FE}{
    short = FE,
    long = Finite Element,
}

\DeclareAcronym{ph}{
    short = PH,
    long = Port-Hamiltonian,
}

\DeclareAcronym{dof}{
    short = DOF,
    long = degree of freedom,
    long-plural-form  = degrees of freedom,
}

\tikzumlset{fill class = white}
\usetikzlibrary{3d,decorations.markings,arrows}
\tikzset{
    MyPersp/.style = {
        scale = 1.0,
        x = {(-0.9cm, 0.3cm)},
        y = {(-0.8cm, -0.4cm)},
        z = {(0cm, 1cm)}
    },
    ->-/.style={decoration={
            markings,
            mark=at position 0.7 with {\arrow[scale=2]{latex}}},postaction={decorate}
    }
}

\makeatletter
\tikzoption{canvas is plane}[]{\@setOxy#1}
\def\@setOxy O(#1,#2,#3)x(#4,#5,#6)y(#7,#8,#9)%
{\def\tikz@plane@origin{\pgfpointxyz{#1}{#2}{#3}}%
    \def\tikz@plane@x{\pgfpointxyz{#4}{#5}{#6}}%
    \def\tikz@plane@y{\pgfpointxyz{#7}{#8}{#9}}%
    \tikz@canvas@is@plane
}
\makeatother

\makenomenclature

\begin{document}
\begin{frontmatter}

This work has been submitted to IFAC for possible publication

\title{An Object-Oriented Library for Heat Transfer Modelling and Simulation in Open Cell Foams\thanksref{footnoteinfo}}

\thanks[footnoteinfo]{This work was supported by 
Deutsche Forschungsgemeinschaft (project number 317092854) 
and Agence Nationale de la Recherche (ID ANR-16-CE92-0028),
project DFG-ANR INFIDHEM.}

\author[First]{Tobias M. Scheuermann} 
\author[First]{Paul Kotyczka} 
\author[First]{Christian Martens} 
\author[Second]{Haithem Louati} 
\author[Second]{Bernhard Maschke} 
\author[Third]{Marie-Line Zanota} 
\author[Second]{Isabelle Pitault}

\address[First]{Technical University of Munich, Department of Mechanical Engineering, Chair of Automatic Control, Boltzmannstraße 15, 85748 Garching, Germany.}
\address[Second]{Univ. Lyon, Université Claude Bernard Lyon 1, CNRS, LAGEPP UMR 5007, 43 boulevard du 11 novembre 1918, F-69100 Villeurbanne, France.}
\address[Third]{Univ. Lyon, CNRS, CPE Lyon, UCBL, LGPC UMR 5285, 43 boulevard du 11 novembre 1918, F-69100 Villeurbanne, France.}

\begin{abstract}                

Metallic open cell foams have multiple applications in industry, 
e.\,g. as catalyst supports in chemical processes. 
Their regular or heterogeneous microscopic structure determines the 
macroscopic thermodynamic and chemical properties. 
We present an object-oriented python library that generates state 
space models for simulation and control from the microscopic foam data, which can be 
imported from the image processing tool iMorph. The foam topology and the 3D geometric data 
are the basis for discrete modeling of the balance laws using the cell method. 
While the material structure imposes a primal chain complex to define discrete 
thermodynamic driving forces, the internal energy balance is evaluated on a second chain complex,
which is constructed by topological duality. The heat exchange between the solid and 
the fluid phase is described based on the available surface data. 
We illustrate in detail the construction of the dual chain complexes, 
and we show how the structured discrete model directly maps to the software objects of
the python code.
As a test case, we present simulation results for a foam with a Kelvin cell structure, 
and compare them to a surrogate finite element model with homogeneous parameters.

\end{abstract}


\end{frontmatter}


\section{Introduction}
Metallic foams are a type of material that is used in multiple ways for industrial purposes.
Two classes of metallic foams are distinguished: closed and open cell foams.
In closed cell foams, the fluid phase is encapsulated in closed cavities inside the foam. 
Open cell foams have connected porous cells so that the fluid can flow through the material.
In this paper, we will concentrate on the latter. 
Due to their high surface to volume ratio, open cell metallic foams are used in catalytic reactors, see e.\,g. \cite{Frey2016Ope}.
In order to design and control the chemical processes in a reactor,
numerical models of the thermodynamic behaviour are needed.
Existing approaches use effective properties, e.\,g. from volume averaging over Cartesian unit cells \citep{Quintard1997Two}.

With the use of tomography, precise 3D voxel data of a given foam sample can be generated
and topological as well as geometric data can be extracted using image processing software like iMorph \citep{brun2008imorph}. 
We will show an approach to set up a numerical model for the heat transfer on open cell foams that is  directly based on the possibly heterogeneous foam topology. Microscopic material parameters and the exact geometry complete the model in the discrete constitutive equations.

The separation of a (Dirac) interconnection structure to describe the structural exchange of power (or the time derivative of another appropriate potential) via pairs of conjugated port variables from material-dependent constitutive equations and energy storage, is at the heart of the \ac{ph} framework, see e.\,g. \cite{Duindam2009Mod} for an overview. In \cite{seslija2014explicit}, and later for non-uniform boundary conditions in \cite{Kotyczka2017Dis}, the discrete modelling of conservation laws on dual chain complexes was presented.
The preliminary work \citep{Scheuermann2019Num} illustrates the discrete modelling of heat transfer and exchange on open cell foams.


In this paper, we adopt this paradigm for the computer-based modelling and simulation of heat transfer on open cell foams. We present the necessary extensions for the classification of topological objects from the regular 2D case as presented in \cite{Kotyczka2017Dis} to irregular 3D meshes in Section 2. The structured representation of the coupled heat equation on dual complexes is presented in Section 3, while we show how this model directly maps to the object oriented python code in Section 4. A numerical example is given in Section 5, and the paper closes with final remarks and an outlook in Section 6.


\section{Image processing}

The input data for model generation and simulation is obtained 
from the image processing tool iMorph%
\footnote{http://imorph.sourceforge.net/}.
iMorph can extract the structure of the foam from 3D tomography pictures.
A typical example is shown below.
Fig. \ref{fig:iMorphFull} shows the image of an open cell foam sample, 
while Fig. \ref{fig:iMorphGraph} displays the extracted solid graph.

\begin{figure}[!ht]
    \centering
    \subfloat[Surface\label{fig:iMorphFull}]
    {
        \includegraphics[width=4cm]{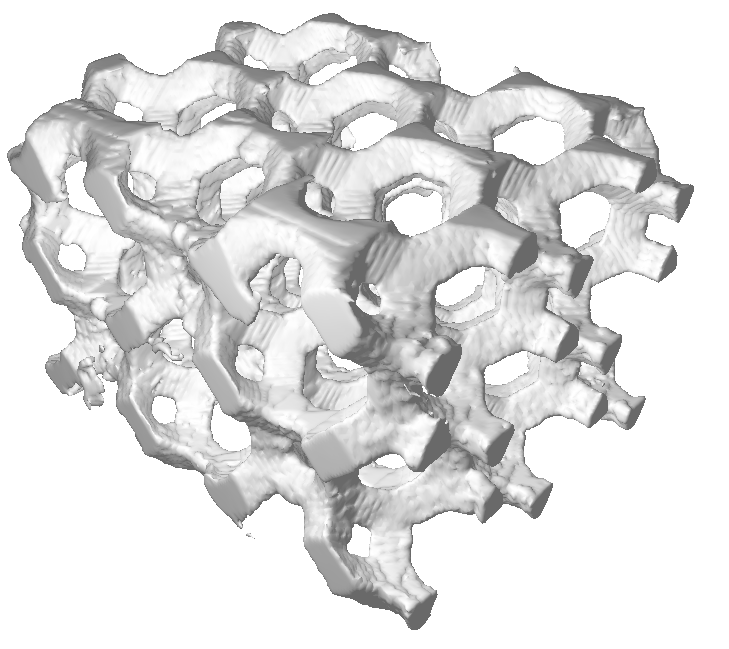}
    }
    \quad
    \subfloat[Graph\label{fig:iMorphGraph}]
    {
        \includegraphics[width=4cm]{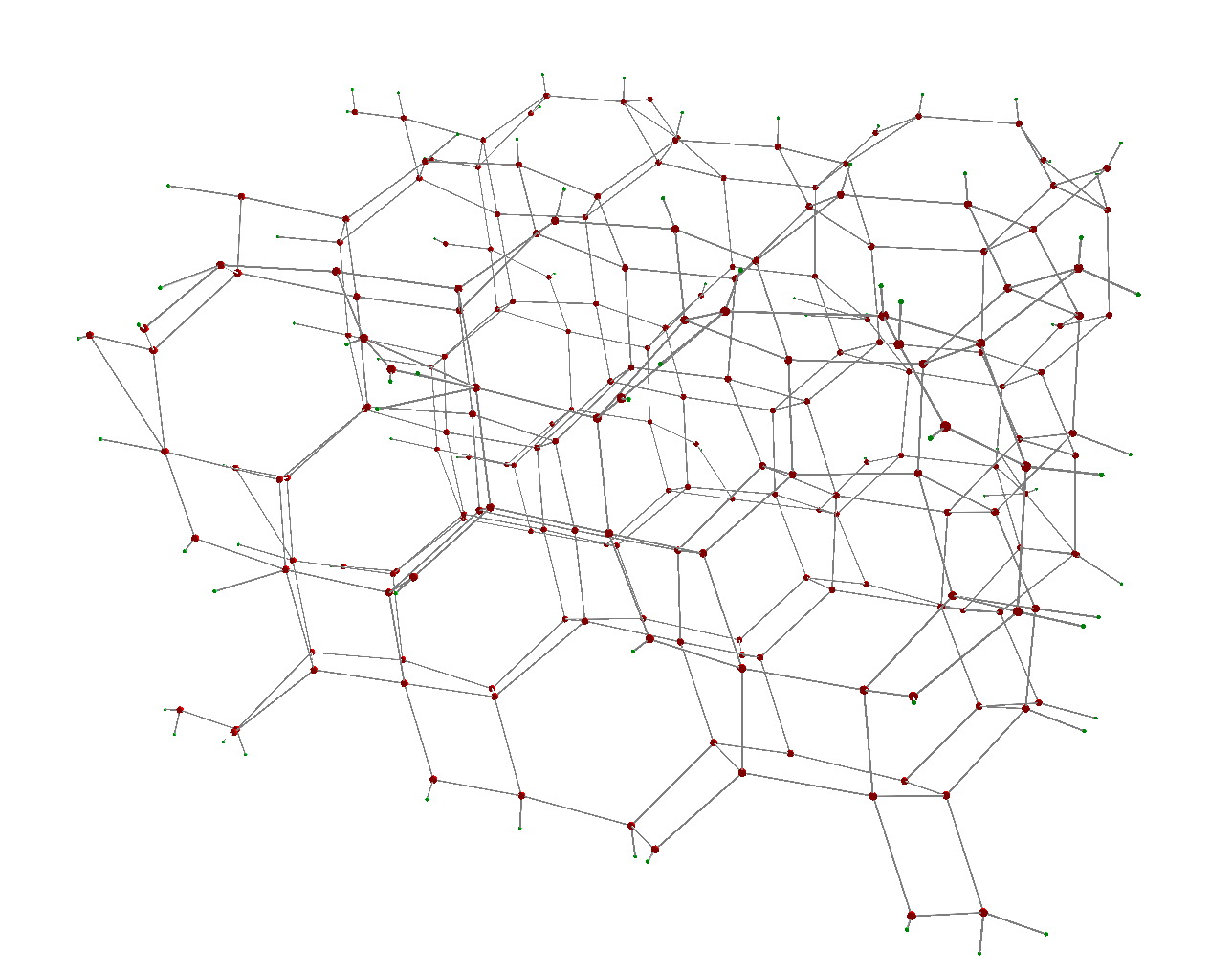}
    }
    \caption{Metallic open cell foam (Kelvin cells)}
       \label{fig:iMorphFoam}
\end{figure}

\begin{figure}[!ht]
    \centering
    \subfloat[Node\label{fig:iMorphNode}]
    {
        \includegraphics[width=3cm]{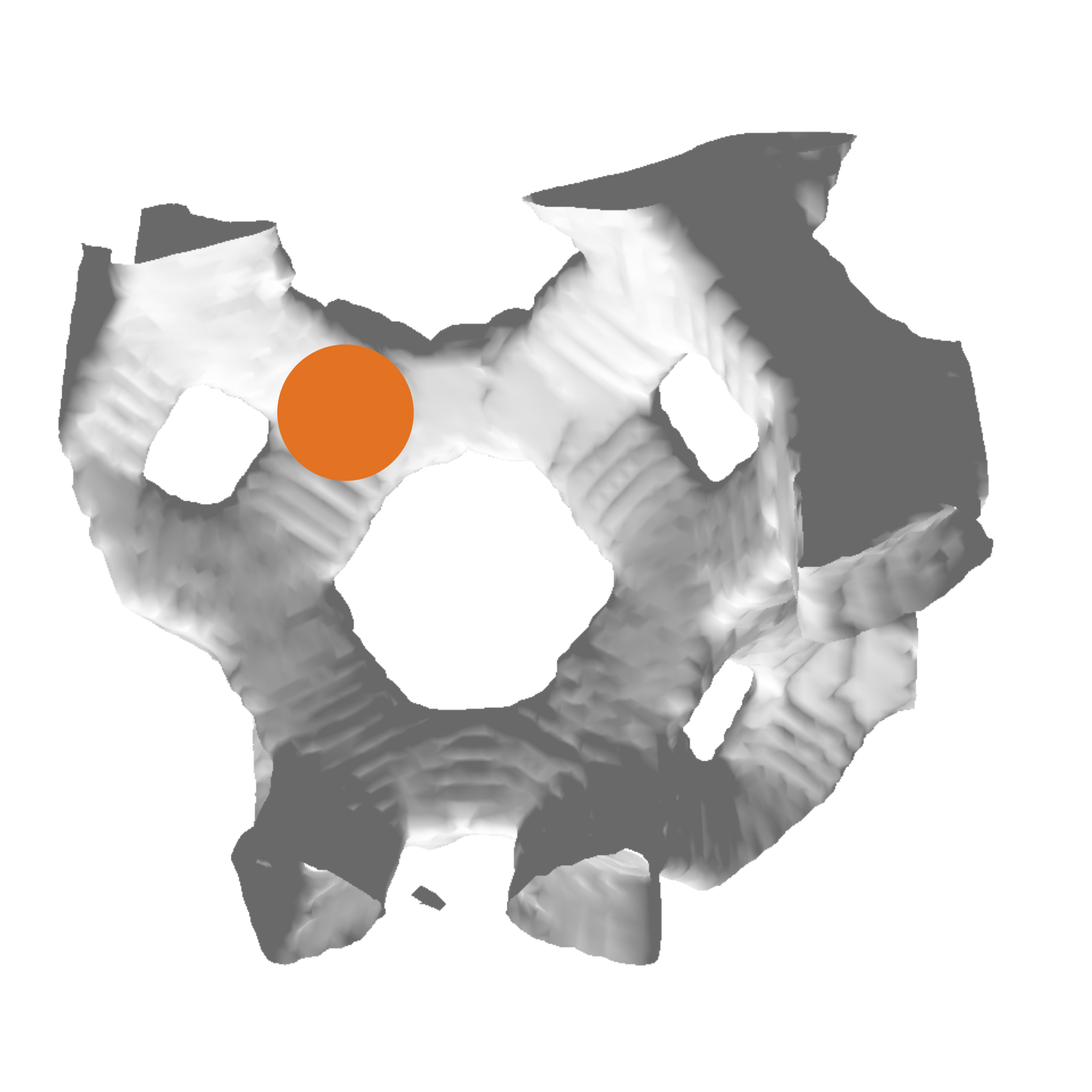}
    }
    \qquad
    \subfloat[Strut\label{fig:iMorphStrut}]
    {
        \includegraphics[width=3cm]{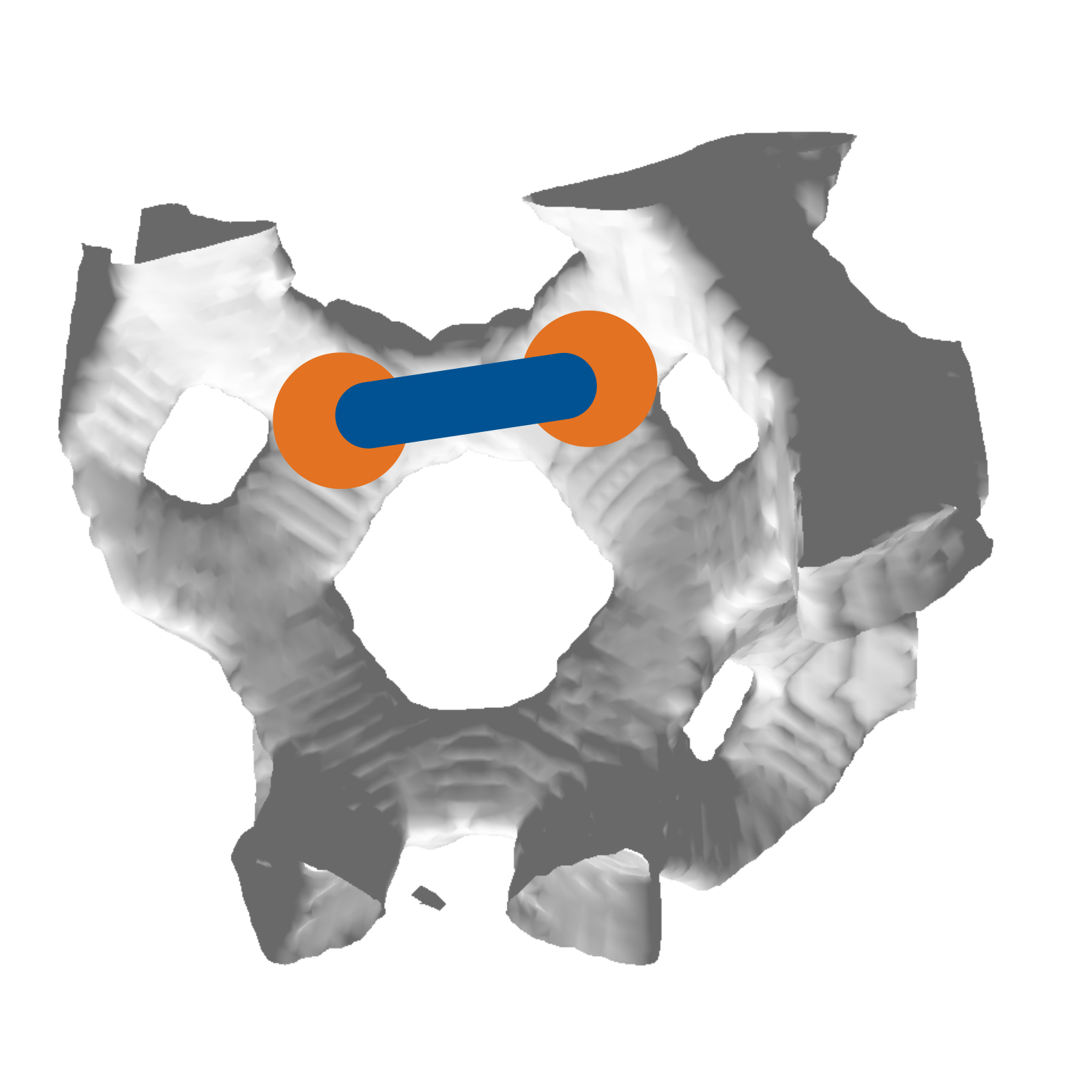}
    }
    \\
    \subfloat[Cell\label{fig:iMorphCell}]
    {
        \includegraphics[width=3cm]{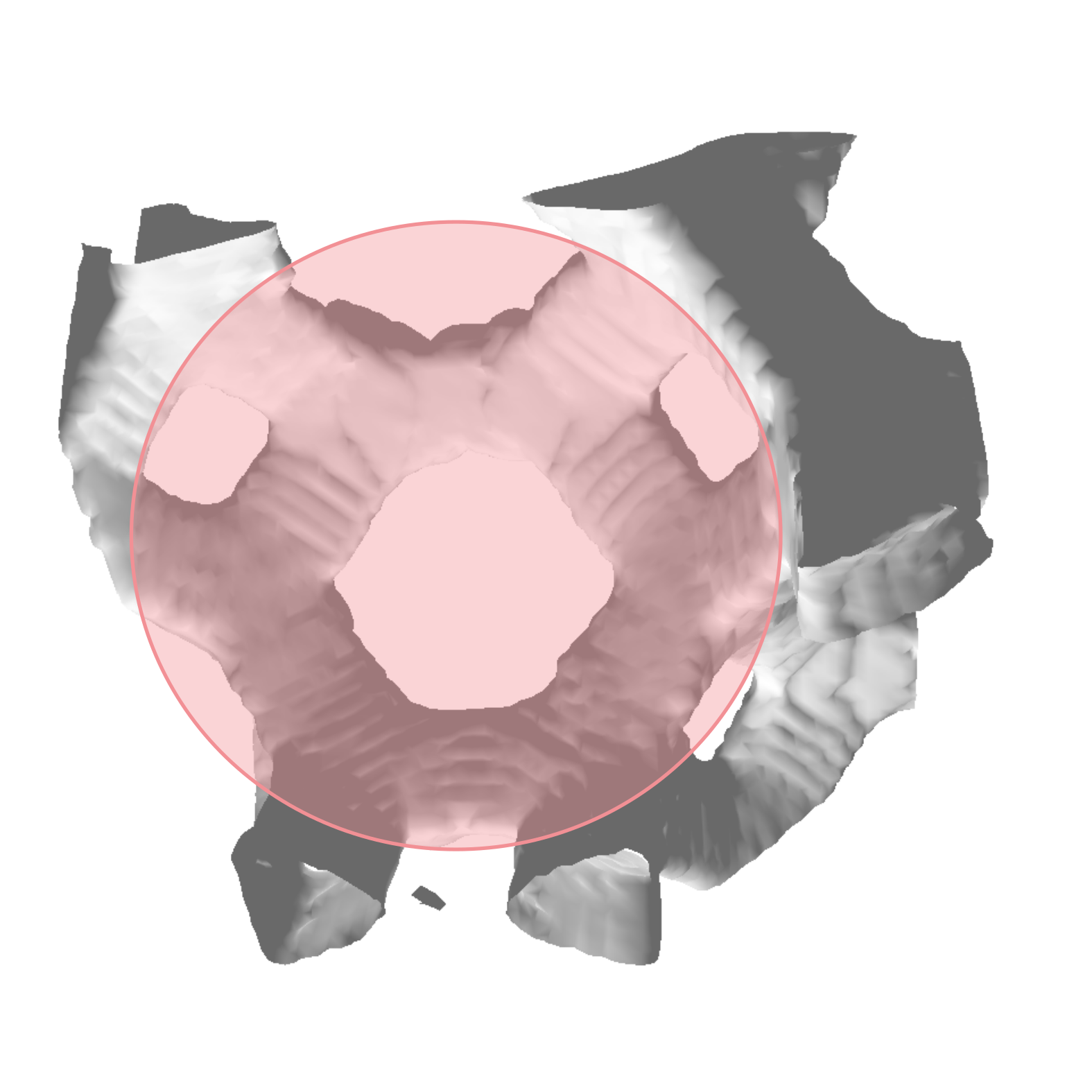}
    }
    \qquad
    \subfloat[Window\label{fig:iMorphWindow}]
    {
        \includegraphics[width=3cm]{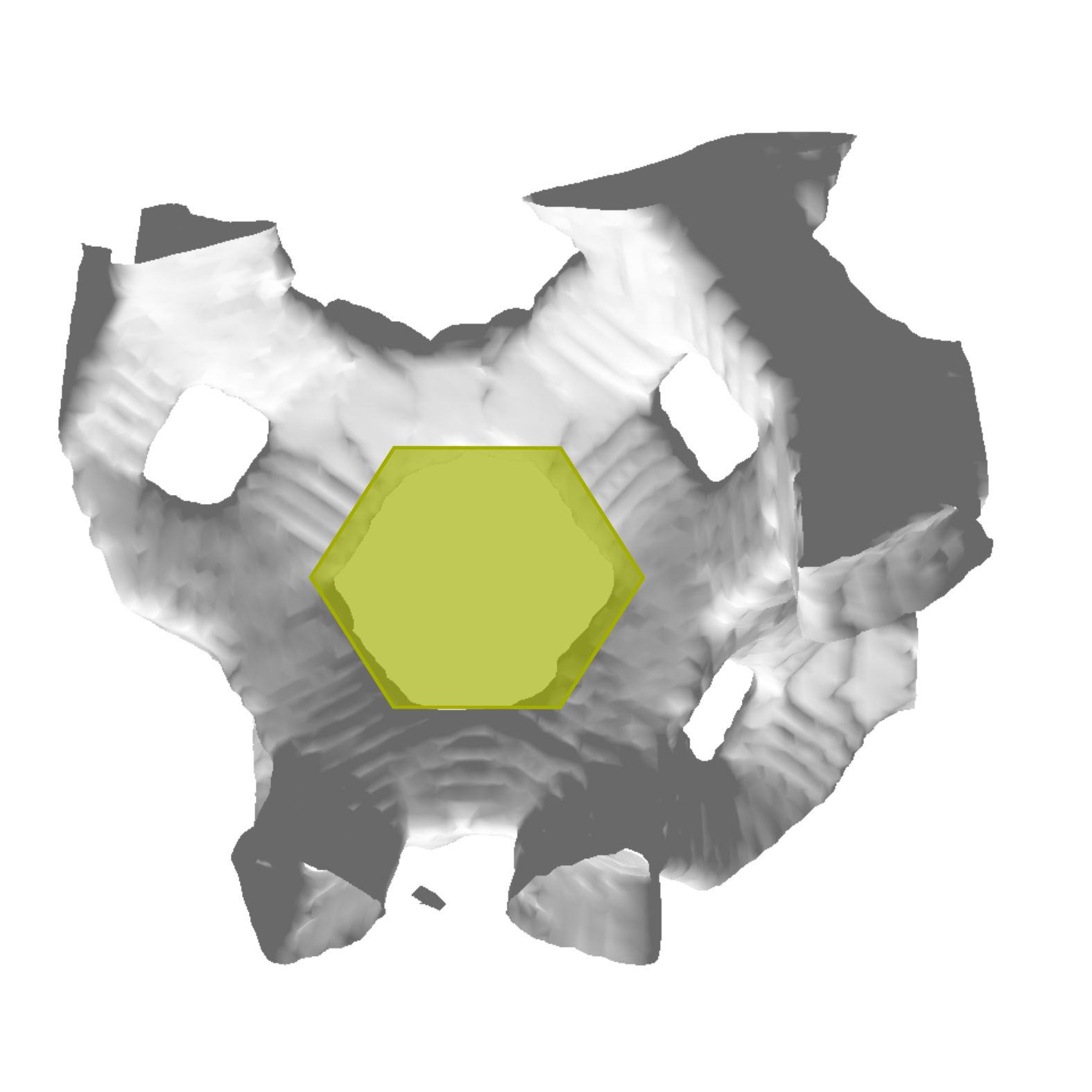}
    }
    \caption{Structures in open cell foams defined by iMorph}
    \label{fig:iMorphStructures}
\end{figure}

Besides the solid nodes (Fig. \ref{fig:iMorphNode}) and struts (Fig. \ref{fig:iMorphStrut}), 
which are represented by the edges of the solid graph, 
iMorph identifies cells (Fig. \ref{fig:iMorphCell}) in the fluid phase.
These cells are connected by so-called ``windows'' (Fig. \ref{fig:iMorphWindow}).

\section{Discrete Heat Equation on Dual Complexes}
The modelling is based on the cell method, 
see \cite{Alotto2013} for an introduction to this numerical 
scheme with references to the original works \citep{Tonti2001ADi} and applications.
The \ac{ph} framework explicitly considers open systems, 
i.\,e. systems with boundary energy flow, see \cite{seslija2014explicit}
for the discrete modelling of conservation laws
and \cite{vanderSchaft2013Por} for \ac{ph} systems on graphs.
We follow the regular 2D approach described in \cite{Kotyczka2017Dis}. 
The heterogeneous 3D case considered here requires some adaptations and additions, 
which are illustrated below.

\subsection{Cells, Chains and Chain Complex}
The topology and geometry of the foam is described in a structured way
using $j$-dimensional cells, or in short ``$j$-cells''\footnote{
The term ``cell'' is used in two contexts, 
that should not be confused with each other: 
It is used in iMorph to describe a cavity in the foam or a $j$-dimensional 
geometric object. 
Therefore, the latter is always denoted as $j$-cell.}, 
see \cite{Arnold1989Mat}, Section 35.D or \cite{Flanders1989Dif}, Section 5.5.
A $j$-cell is a geometric object that consists of a convex polyhedron $D \subset \mathbb{R}^j$, 
a differentiable $f: D \rightarrow M$ on the $n$-dimensional manifold $M$
and an orientation.
A formal sum of $j$-cells is called $j$-chain.

The linear vector space of $j$-chains on a tessellation $K$ is denoted $C_j(K,\mathbb{R})$.
The boundary of each $j$-cell consists of a $j-1$-chain 
and is found by applying the boundary operator $\partial_j$.
Applying the boundary operator twice to a $j$-chain results in an empty set, 
which is the central property of a chain complex, see e.\,g. \cite{jaenich2001vector},
 Section 7.6. The spaces of $j$-chains, $j=n,\ldots,0$, which,
 connected via the boundary operators, form an $n$-complex, 
can be represented in a sequence diagram:

\begin{align}
    C_n(K,\mathbb{R}) \overset{\partial_n}{\longrightarrow} 
    C_{n-1}(K,\mathbb{R}) \overset{\partial_{n-1}}{\longrightarrow} \ldots
    \overset{\partial_1}{\longrightarrow} C_0(K,\mathbb{R}) 
\end{align}

In the following, we call a $n$-chain with the collection
of all $j$-cells, $j=0...n$, appearing in the sequence above, an $n$-complex.
The symbol $\partial$ will be used for both the boundary operator and its matrix representation,
i.\,e. an incidence matrix.
For our application, only the case with $n=3$ is relevant, 
so we will restrict ourselves to this case.

\subsection{Definition of the Primal 3-Complex}
The primal 3-complex is initially given by the structure of the solid phase. 
Since an $n$-complex can be seen as a generalized \emph{directed} graph,
orientations have to be assigned to all $j$-cells.
The nodes (0-cells) and edges (1-cells) of the primal 3-complex can be taken directly from the graph generated with iMorph.
Faces (2-cells) correspond to the iMorph windows. The windows that enclose a fluid cell 
define a volume (3-cells).

\begin{figure}[!ht]
    \centering
    \begin{tikzpicture}[MyPersp,scale=1.8]
\node[circle,fill,TUMOrange] (nb0) at (0.0,0.0,0.0) {};
\node[circle,fill,TUMOrange] (nb1) at (1.0,0.0,0.0) {};
\node[circle,fill,TUMOrange] (nb2) at (0.0,1.0,0.0) {};
\node[circle,fill,TUMOrange] (nb3) at (1.0,1.0,0.0) {};
\node[circle,fill,TUMOrange] (nb4) at (0.0,0.0,1.0) {};
\node[circle,fill,TUMOrange] (nb5) at (1.0,0.0,1.0) {};
\node[circle,fill,TUMOrange] (nb6) at (0.0,1.0,1.0) {};
\node[circle,fill,TUMOrange] (ni0) at (1.0,1.0,1.0) {};
\node[circle,draw,TUMOrange] (nB0) at (1.5,1.5,1.5) {};
\coordinate (n0) at (1.5,0.0,0.0) ;
\coordinate (n1) at (0.0,1.5,0.0) ;
\coordinate (n2) at (0.0,0.0,1.5) ;
\coordinate (n3) at (1.5,1.5,0.0) ;
\coordinate (n4) at (1.5,0.0,1.5) ;
\coordinate (n5) at (0.0,1.5,1.5) ;
\draw[->-,TUMBlue] (nb0) -- (nb1);
\draw[->-,TUMBlue] (nb1) -- (nb3);
\draw[->-,TUMBlue] (nb3) -- (nb2);
\draw[->-,TUMBlue] (nb2) -- (nb0);
\draw[->-,TUMBlue] (nb4) -- (nb5);
\draw[->-,TUMBlue] (nb5) -- (ni0);
\draw[->-,TUMBlue] (ni0) -- (nb6);
\draw[->-,TUMBlue] (nb6) -- (nb4);
\draw[->-,TUMBlue] (nb0) -- (nb4);
\draw[->-,TUMBlue] (nb1) -- (nb5);
\draw[->-,TUMBlue] (nb2) -- (nb6);
\draw[->-,TUMBlue] (nb3) -- (ni0);
\draw[->-,TUMBlue] (ni0) -- (nB0);
\draw[->-,TUMBlue] (nB0) -- (n4);
\draw[->-,TUMBlue] (n4) -- (nb5);
\draw[->-,TUMBlue] (nB0) -- (n3);
\draw[->-,TUMBlue] (n3) -- (nb3);
\draw[->-,TUMBlue] (nb6) -- (n5);
\draw[->-,TUMBlue] (n5) -- (nB0);
\draw[dashed,TUMBlue] (nb1) -- (n0);
\draw[dashed,TUMBlue] (n0) -- (n3);
\draw[dashed,TUMBlue] (nb2) -- (n1);
\draw[dashed,TUMBlue] (n1) -- (n3);
\draw[dashed,TUMBlue] (nb4) -- (n2);
\draw[dashed,TUMBlue] (n0) -- (n4);
\draw[dashed,TUMBlue] (n1) -- (n5);
\draw[dashed,TUMBlue] (n5) -- (n2);
\draw[dashed,TUMBlue] (n2) -- (n4);
\begin{scope}[canvas is plane={O(0.5,0.5,0.0)x(1.5,0.5,0.0)y(0.5,1.5,0.0)}]
	\node[TUMGreen] (fb0) at (0,0) {};
	\circledarrow{TUMGreen}{fb0}{0.2}
\end{scope}
\fill[TUMGreen,fill opacity=0.05] (nb0.center) -- (nb1.center) -- (nb3.center) -- (nb2.center) -- cycle;
\begin{scope}[canvas is plane={O(1.0,0.5,0.5)x(1.0,1.5,0.5)y(1.0,0.5,1.5)}]
	\node[TUMGreen] (fi0) at (0,0) {};
	\circledarrow{TUMGreen}{fi0}{0.2}
\end{scope}
\fill[TUMGreen,fill opacity=0.05] (nb1.center) -- (nb3.center) -- (ni0.center) -- (nb5.center) -- cycle;
\begin{scope}[canvas is plane={O(0.5,1.0,0.5)x(1.5,1.0,0.5)y(0.5,1.0,-0.5)}]
	\node[TUMGreen] (fi1) at (0,0) {};
	\circledarrow{TUMGreen}{fi1}{0.2}
\end{scope}
\fill[TUMGreen,fill opacity=0.05] (nb3.center) -- (nb2.center) -- (nb6.center) -- (ni0.center) -- cycle;
\begin{scope}[canvas is plane={O(0.0,0.5,0.5)x(0.0,1.5,0.5)y(0.0,0.5,-0.5)}]
	\node[TUMGreen] (fb1) at (0,0) {};
	\circledarrow{TUMGreen}{fb1}{0.2}
\end{scope}
\fill[TUMGreen,fill opacity=0.05] (nb2.center) -- (nb0.center) -- (nb4.center) -- (nb6.center) -- cycle;
\begin{scope}[canvas is plane={O(0.5,0.0,0.5)x(1.5,0.0,0.5)y(0.5,0.0,1.5)}]
	\node[TUMGreen] (fb2) at (0,0) {};
	\circledarrow{TUMGreen}{fb2}{0.2}
\end{scope}
\fill[TUMGreen,fill opacity=0.05] (nb0.center) -- (nb1.center) -- (nb5.center) -- (nb4.center) -- cycle;
\begin{scope}[canvas is plane={O(0.5,0.5,1.0)x(1.5,0.5,1.0)y(0.5,1.5,1.0)}]
	\node[TUMGreen] (fi2) at (0,0) {};
	\circledarrow{TUMGreen}{fi2}{0.2}
\end{scope}
\fill[TUMGreen,fill opacity=0.05] (nb4.center) -- (nb5.center) -- (ni0.center) -- (nb6.center) -- cycle;
\begin{scope}[canvas is plane={O(1.27,0.633,0.0)x(2.27,0.633,0.0)y(1.27,1.63,0.0)}]
	\node[TUMGreen] (fB0a) at (0,0) {};
	\circledarrow{TUMGreen}{fB0a}{0.2}
\end{scope}
\fill[TUMGreen,fill opacity=0.05] (nb1.center) -- (n0.center) -- (n3.center) -- (nb3.center) -- cycle;
\begin{scope}[canvas is plane={O(1.5,0.75,0.75)x(1.5,1.75,0.75)y(1.5,0.75,-0.25)}]
	\node[TUMGreen] (fB0b) at (0,0) {};
	\circledarrow{TUMGreen}{fB0b}{0.2}
\end{scope}
\fill[TUMGreen,fill opacity=0.05] (n0.center) -- (n4.center) -- (nB0.center) -- (n3.center) -- cycle;
\begin{scope}[canvas is plane={O(1.27,0.0,0.633)x(2.27,0.0,0.633)y(1.27,0.0,-0.367)}]
	\node[TUMGreen] (fB0c) at (0,0) {};
	\circledarrow{TUMGreen}{fB0c}{0.2}
\end{scope}
\fill[TUMGreen,fill opacity=0.05] (nb1.center) -- (nb5.center) -- (n4.center) -- (n0.center) -- cycle;
\begin{scope}[canvas is plane={O(1.27,0.633,1.27)x(1.27,1.63,1.27)y(1.97,0.633,1.97)}]
	\node[TUMGreen] (fi3) at (0,0) {};
	\circledarrow{TUMGreen}{fi3}{0.2}
\end{scope}
\fill[TUMGreen,fill opacity=0.05] (nb5.center) -- (ni0.center) -- (nB0.center) -- (n4.center) -- cycle;
\begin{scope}[canvas is plane={O(1.27,1.27,0.633)x(1.27,1.27,1.63)y(1.97,1.97,0.633)}]
	\node[TUMGreen] (fi4) at (0,0) {};
	\circledarrow{TUMGreen}{fi4}{0.2}
\end{scope}
\fill[TUMGreen,fill opacity=0.05] (nb3.center) -- (ni0.center) -- (nB0.center) -- (n3.center) -- cycle;
\begin{scope}[canvas is plane={O(0.633,1.27,0.0)x(1.63,1.27,0.0)y(0.633,0.267,0.0)}]
	\node[TUMGreen] (fB1a) at (0,0) {};
	\circledarrow{TUMGreen}{fB1a}{0.2}
\end{scope}
\fill[TUMGreen,fill opacity=0.05] (nb3.center) -- (nb2.center) -- (n1.center) -- (n3.center) -- cycle;
\begin{scope}[canvas is plane={O(0.75,1.5,0.75)x(1.75,1.5,0.75)y(0.75,1.5,-0.25)}]
	\node[TUMGreen] (fB1b) at (0,0) {};
	\circledarrow{TUMGreen}{fB1b}{0.2}
\end{scope}
\fill[TUMGreen,fill opacity=0.05] (n1.center) -- (n5.center) -- (nB0.center) -- (n3.center) -- cycle;
\begin{scope}[canvas is plane={O(0.0,1.27,0.633)x(0.0,2.27,0.633)y(0.0,1.27,-0.367)}]
	\node[TUMGreen] (fB1c) at (0,0) {};
	\circledarrow{TUMGreen}{fB1c}{0.2}
\end{scope}
\fill[TUMGreen,fill opacity=0.05] (nb2.center) -- (nb6.center) -- (n5.center) -- (n1.center) -- cycle;
\begin{scope}[canvas is plane={O(0.633,1.27,1.27)x(1.63,1.27,1.27)y(0.633,0.56,0.56)}]
	\node[TUMGreen] (fi5) at (0,0) {};
	\circledarrow{TUMGreen}{fi5}{0.2}
\end{scope}
\fill[TUMGreen,fill opacity=0.05] (ni0.center) -- (nb6.center) -- (n5.center) -- (nB0.center) -- cycle;
\begin{scope}[canvas is plane={O(0.633,0.0,1.27)x(1.63,0.0,1.27)y(0.633,0.0,2.27)}]
	\node[TUMGreen] (fB2a) at (0,0) {};
	\circledarrow{TUMGreen}{fB2a}{0.2}
\end{scope}
\fill[TUMGreen,fill opacity=0.05] (nb4.center) -- (nb5.center) -- (n4.center) -- (n2.center) -- cycle;
\begin{scope}[canvas is plane={O(0.75,0.75,1.5)x(1.75,0.75,1.5)y(0.75,1.75,1.5)}]
	\node[TUMGreen] (fB2b) at (0,0) {};
	\circledarrow{TUMGreen}{fB2b}{0.2}
\end{scope}
\fill[TUMGreen,fill opacity=0.05] (n2.center) -- (n4.center) -- (nB0.center) -- (n5.center) -- cycle;
\begin{scope}[canvas is plane={O(0.0,0.633,1.27)x(0.0,1.63,1.27)y(0.0,0.633,0.267)}]
	\node[TUMGreen] (fB2c) at (0,0) {};
	\circledarrow{TUMGreen}{fB2c}{0.2}
\end{scope}
\fill[TUMGreen,fill opacity=0.05] (nb6.center) -- (nb4.center) -- (n2.center) -- (n5.center) -- cycle;
\end{tikzpicture}
    \caption{Primal complex}
    \label{fig:primalComplex}
\end{figure}
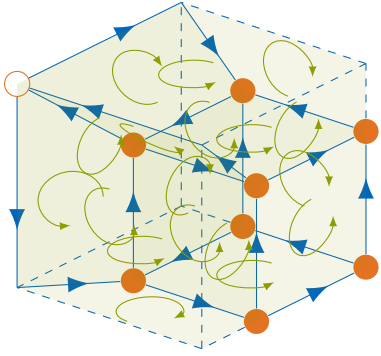

The following classification of inner and border $j$-cells 
is necessary for the direct imposition of boundary conditions in the 
numerical model.
To realize \acp{nbc}, 
i.\,e. heat flux boundary conditions on the appropriate dual objects, 
see Subsection \ref{subsec:dual-complex}, 
additional border nodes must be defined, 
which lead to additional edges, 
faces and volumes on a thin, 
artificial boundary layer.

Fig. \ref{fig:primalComplex} shows a minimal example 
for a $3$-complex with the orientation of the $j$-cells. 
The different categories of $j$-cells are described below.

\paragraph*{Inner nodes:} 
Solid nodes inside the domain are 
called inner nodes and are denoted by $n_\mathrm{i}\in\mathcal{N}_\mathrm{i}$.

\paragraph*{Border nodes:}
Solid nodes on the boundary are called 
border nodes and are denoted by $n_\mathrm{b}\in\mathcal{N}_\mathrm{b}$. 
At these nodes, a \ac{dbc} is imposed.

\paragraph*{Additional border nodes:}
These nodes, denoted by $n_\mathrm{b}\in\mathcal{N}_\mathrm{B}$, 
are not a representation of a solid node, 
but an intersection of a strut with the boundary (border edge, see below). 
Through the dual face to this edge, see next subsection, a \ac{nbc} is imposed.

\begin{figure}[!ht]
    \centering
    \subfloat[Inner\label{fig:primalInnerNodes}]
    {
        \begin{tikzpicture}[MyPersp,scale=0.8]
\coordinate (nb0) at (0.0,0.0,0.0) ;
\coordinate (nb1) at (1.0,0.0,0.0) ;
\coordinate (nb2) at (0.0,1.0,0.0) ;
\coordinate (nb3) at (1.0,1.0,0.0) ;
\coordinate (nb4) at (0.0,0.0,1.0) ;
\coordinate (nb5) at (1.0,0.0,1.0) ;
\coordinate (nb6) at (0.0,1.0,1.0) ;
\node[circle,fill,inner sep=0pt, minimum size=2mm,TUMOrange] (ni0) at (1.0,1.0,1.0) {};
\coordinate (nB0) at (1.5,1.5,1.5) ;
\coordinate (n0) at (1.5,0.0,0.0) ;
\coordinate (n1) at (0.0,1.5,0.0) ;
\coordinate (n2) at (0.0,0.0,1.5) ;
\coordinate (n3) at (1.5,1.5,0.0) ;
\coordinate (n4) at (1.5,0.0,1.5) ;
\coordinate (n5) at (0.0,1.5,1.5) ;
\draw[TUMGrayMedium] (nb0) -- (nb1);
\draw[TUMGrayMedium] (nb1) -- (nb3);
\draw[TUMGrayMedium] (nb3) -- (nb2);
\draw[TUMGrayMedium] (nb2) -- (nb0);
\draw[TUMGrayMedium] (nb4) -- (nb5);
\draw[TUMGrayMedium] (nb5) -- (ni0);
\draw[TUMGrayMedium] (ni0) -- (nb6);
\draw[TUMGrayMedium] (nb6) -- (nb4);
\draw[TUMGrayMedium] (nb0) -- (nb4);
\draw[TUMGrayMedium] (nb1) -- (nb5);
\draw[TUMGrayMedium] (nb2) -- (nb6);
\draw[TUMGrayMedium] (nb3) -- (ni0);
\draw[TUMGrayMedium] (ni0) -- (nB0);
\draw[TUMGrayMedium] (nB0) -- (n4);
\draw[TUMGrayMedium] (n4) -- (nb5);
\draw[TUMGrayMedium] (nB0) -- (n3);
\draw[TUMGrayMedium] (n3) -- (nb3);
\draw[TUMGrayMedium] (nb6) -- (n5);
\draw[TUMGrayMedium] (n5) -- (nB0);
\draw[dashed,TUMGrayMedium] (nb1) -- (n0);
\draw[dashed,TUMGrayMedium] (n0) -- (n3);
\draw[dashed,TUMGrayMedium] (nb2) -- (n1);
\draw[dashed,TUMGrayMedium] (n1) -- (n3);
\draw[dashed,TUMGrayMedium] (nb4) -- (n2);
\draw[dashed,TUMGrayMedium] (n0) -- (n4);
\draw[dashed,TUMGrayMedium] (n1) -- (n5);
\draw[dashed,TUMGrayMedium] (n5) -- (n2);
\draw[dashed,TUMGrayMedium] (n2) -- (n4);
\end{tikzpicture}
    }
    \quad
    \subfloat[Border\label{fig:primalBorderNodes}]
    {
        \begin{tikzpicture}[MyPersp,scale=0.8]
\node[circle,fill,inner sep=0pt, minimum size=2mm,TUMOrange] (nb0) at (0.0,0.0,0.0) {};
\node[circle,fill,inner sep=0pt, minimum size=2mm,TUMOrange] (nb1) at (1.0,0.0,0.0) {};
\node[circle,fill,inner sep=0pt, minimum size=2mm,TUMOrange] (nb2) at (0.0,1.0,0.0) {};
\node[circle,fill,inner sep=0pt, minimum size=2mm,TUMOrange] (nb3) at (1.0,1.0,0.0) {};
\node[circle,fill,inner sep=0pt, minimum size=2mm,TUMOrange] (nb4) at (0.0,0.0,1.0) {};
\node[circle,fill,inner sep=0pt, minimum size=2mm,TUMOrange] (nb5) at (1.0,0.0,1.0) {};
\node[circle,fill,inner sep=0pt, minimum size=2mm,TUMOrange] (nb6) at (0.0,1.0,1.0) {};
\coordinate (ni0) at (1.0,1.0,1.0) ;
\coordinate (nB0) at (1.5,1.5,1.5) ;
\coordinate (n0) at (1.5,0.0,0.0) ;
\coordinate (n1) at (0.0,1.5,0.0) ;
\coordinate (n2) at (0.0,0.0,1.5) ;
\coordinate (n3) at (1.5,1.5,0.0) ;
\coordinate (n4) at (1.5,0.0,1.5) ;
\coordinate (n5) at (0.0,1.5,1.5) ;
\draw[TUMGrayMedium] (nb0) -- (nb1);
\draw[TUMGrayMedium] (nb1) -- (nb3);
\draw[TUMGrayMedium] (nb3) -- (nb2);
\draw[TUMGrayMedium] (nb2) -- (nb0);
\draw[TUMGrayMedium] (nb4) -- (nb5);
\draw[TUMGrayMedium] (nb5) -- (ni0);
\draw[TUMGrayMedium] (ni0) -- (nb6);
\draw[TUMGrayMedium] (nb6) -- (nb4);
\draw[TUMGrayMedium] (nb0) -- (nb4);
\draw[TUMGrayMedium] (nb1) -- (nb5);
\draw[TUMGrayMedium] (nb2) -- (nb6);
\draw[TUMGrayMedium] (nb3) -- (ni0);
\draw[TUMGrayMedium] (ni0) -- (nB0);
\draw[TUMGrayMedium] (nB0) -- (n4);
\draw[TUMGrayMedium] (n4) -- (nb5);
\draw[TUMGrayMedium] (nB0) -- (n3);
\draw[TUMGrayMedium] (n3) -- (nb3);
\draw[TUMGrayMedium] (nb6) -- (n5);
\draw[TUMGrayMedium] (n5) -- (nB0);
\draw[dashed,TUMGrayMedium] (nb1) -- (n0);
\draw[dashed,TUMGrayMedium] (n0) -- (n3);
\draw[dashed,TUMGrayMedium] (nb2) -- (n1);
\draw[dashed,TUMGrayMedium] (n1) -- (n3);
\draw[dashed,TUMGrayMedium] (nb4) -- (n2);
\draw[dashed,TUMGrayMedium] (n0) -- (n4);
\draw[dashed,TUMGrayMedium] (n1) -- (n5);
\draw[dashed,TUMGrayMedium] (n5) -- (n2);
\draw[dashed,TUMGrayMedium] (n2) -- (n4);
\end{tikzpicture}
    }
    \quad
    \subfloat[Additional border\label{fig:primalAddBorderNodes}]
    {
        \begin{tikzpicture}[MyPersp,scale=0.8]
\coordinate (nb0) at (0.0,0.0,0.0) ;
\coordinate (nb1) at (1.0,0.0,0.0) ;
\coordinate (nb2) at (0.0,1.0,0.0) ;
\coordinate (nb3) at (1.0,1.0,0.0) ;
\coordinate (nb4) at (0.0,0.0,1.0) ;
\coordinate (nb5) at (1.0,0.0,1.0) ;
\coordinate (nb6) at (0.0,1.0,1.0) ;
\coordinate (ni0) at (1.0,1.0,1.0) ;
\node[circle,draw,inner sep=0pt, minimum size=2mm,TUMOrange] (nB0) at (1.5,1.5,1.5) {};
\coordinate (n0) at (1.5,0.0,0.0) ;
\coordinate (n1) at (0.0,1.5,0.0) ;
\coordinate (n2) at (0.0,0.0,1.5) ;
\coordinate (n3) at (1.5,1.5,0.0) ;
\coordinate (n4) at (1.5,0.0,1.5) ;
\coordinate (n5) at (0.0,1.5,1.5) ;
\draw[TUMGrayMedium] (nb0) -- (nb1);
\draw[TUMGrayMedium] (nb1) -- (nb3);
\draw[TUMGrayMedium] (nb3) -- (nb2);
\draw[TUMGrayMedium] (nb2) -- (nb0);
\draw[TUMGrayMedium] (nb4) -- (nb5);
\draw[TUMGrayMedium] (nb5) -- (ni0);
\draw[TUMGrayMedium] (ni0) -- (nb6);
\draw[TUMGrayMedium] (nb6) -- (nb4);
\draw[TUMGrayMedium] (nb0) -- (nb4);
\draw[TUMGrayMedium] (nb1) -- (nb5);
\draw[TUMGrayMedium] (nb2) -- (nb6);
\draw[TUMGrayMedium] (nb3) -- (ni0);
\draw[TUMGrayMedium] (ni0) -- (nB0);
\draw[TUMGrayMedium] (nB0) -- (n4);
\draw[TUMGrayMedium] (n4) -- (nb5);
\draw[TUMGrayMedium] (nB0) -- (n3);
\draw[TUMGrayMedium] (n3) -- (nb3);
\draw[TUMGrayMedium] (nb6) -- (n5);
\draw[TUMGrayMedium] (n5) -- (nB0);
\draw[dashed,TUMGrayMedium] (nb1) -- (n0);
\draw[dashed,TUMGrayMedium] (n0) -- (n3);
\draw[dashed,TUMGrayMedium] (nb2) -- (n1);
\draw[dashed,TUMGrayMedium] (n1) -- (n3);
\draw[dashed,TUMGrayMedium] (nb4) -- (n2);
\draw[dashed,TUMGrayMedium] (n0) -- (n4);
\draw[dashed,TUMGrayMedium] (n1) -- (n5);
\draw[dashed,TUMGrayMedium] (n5) -- (n2);
\draw[dashed,TUMGrayMedium] (n2) -- (n4);
\end{tikzpicture}
    }
    \caption{Primal nodes}
    \label{fig:primalNodes}
\end{figure}
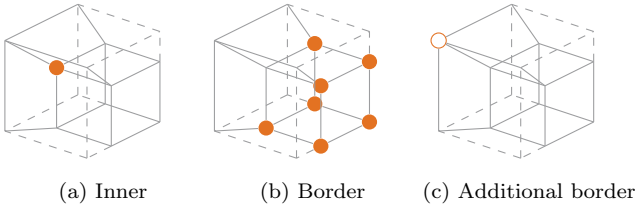

\paragraph*{Inner edges:} 
Inner edges $e_\mathrm{i}\in\mathcal{E}_\mathrm{i}$ are connections of the inner 
nodes and border nodes.
They represent struts that are entirely 
inside the domain or on its boundary 
($n_\mathrm{i}$ with $n_\mathrm{i}$, $n_\mathrm{b}$ with $n_\mathrm{b}$ 
and $n_\mathrm{i}$ with $n_\mathrm{b}$).

\paragraph*{Border edges:} 
Border edges $e_\mathrm{b}\in\mathcal{E}_\mathrm{b}$ connect inner nodes to additional border nodes 
an represent struts that cross the system boundary.

\paragraph*{Additional border edges:} 
These edges $e_\mathrm{B}\in\mathcal{E}_\mathrm{B}$ have no representation in the solid graph, 
but they are necessery to fill the entire domain with volumes.

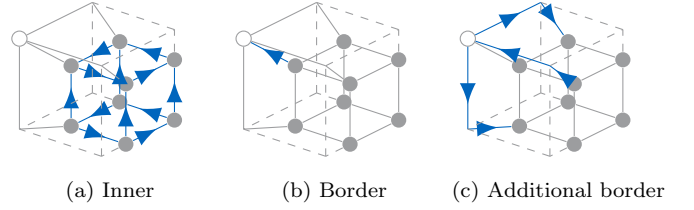
\begin{figure}[!ht]
    \centering
    \subfloat[Inner\label{fig:primalInnerEdges}]
    {
        \begin{tikzpicture}[MyPersp,scale=0.8]
\node[circle,fill,inner sep=0pt, minimum size=2mm,TUMGrayMedium] (nb0) at (0.0,0.0,0.0) {};
\node[circle,fill,inner sep=0pt, minimum size=2mm,TUMGrayMedium] (nb1) at (1.0,0.0,0.0) {};
\node[circle,fill,inner sep=0pt, minimum size=2mm,TUMGrayMedium] (nb2) at (0.0,1.0,0.0) {};
\node[circle,fill,inner sep=0pt, minimum size=2mm,TUMGrayMedium] (nb3) at (1.0,1.0,0.0) {};
\node[circle,fill,inner sep=0pt, minimum size=2mm,TUMGrayMedium] (nb4) at (0.0,0.0,1.0) {};
\node[circle,fill,inner sep=0pt, minimum size=2mm,TUMGrayMedium] (nb5) at (1.0,0.0,1.0) {};
\node[circle,fill,inner sep=0pt, minimum size=2mm,TUMGrayMedium] (nb6) at (0.0,1.0,1.0) {};
\node[circle,fill,inner sep=0pt, minimum size=2mm,TUMGrayMedium] (ni0) at (1.0,1.0,1.0) {};
\node[circle,draw,inner sep=0pt, minimum size=2mm,TUMGrayMedium] (nB0) at (1.5,1.5,1.5) {};
\coordinate (n0) at (1.5,0.0,0.0) ;
\coordinate (n1) at (0.0,1.5,0.0) ;
\coordinate (n2) at (0.0,0.0,1.5) ;
\coordinate (n3) at (1.5,1.5,0.0) ;
\coordinate (n4) at (1.5,0.0,1.5) ;
\coordinate (n5) at (0.0,1.5,1.5) ;
\draw[->-,TUMBlue] (nb0) -- (nb1);
\draw[->-,TUMBlue] (nb1) -- (nb3);
\draw[->-,TUMBlue] (nb3) -- (nb2);
\draw[->-,TUMBlue] (nb2) -- (nb0);
\draw[->-,TUMBlue] (nb4) -- (nb5);
\draw[->-,TUMBlue] (nb5) -- (ni0);
\draw[->-,TUMBlue] (ni0) -- (nb6);
\draw[->-,TUMBlue] (nb6) -- (nb4);
\draw[->-,TUMBlue] (nb0) -- (nb4);
\draw[->-,TUMBlue] (nb1) -- (nb5);
\draw[->-,TUMBlue] (nb2) -- (nb6);
\draw[->-,TUMBlue] (nb3) -- (ni0);
\draw[TUMGrayMedium] (ni0) -- (nB0);
\draw[TUMGrayMedium] (nB0) -- (n4);
\draw[TUMGrayMedium] (n4) -- (nb5);
\draw[TUMGrayMedium] (nB0) -- (n3);
\draw[TUMGrayMedium] (n3) -- (nb3);
\draw[TUMGrayMedium] (nb6) -- (n5);
\draw[TUMGrayMedium] (n5) -- (nB0);
\draw[dashed,TUMGrayMedium] (nb1) -- (n0);
\draw[dashed,TUMGrayMedium] (n0) -- (n3);
\draw[dashed,TUMGrayMedium] (nb2) -- (n1);
\draw[dashed,TUMGrayMedium] (n1) -- (n3);
\draw[dashed,TUMGrayMedium] (nb4) -- (n2);
\draw[dashed,TUMGrayMedium] (n0) -- (n4);
\draw[dashed,TUMGrayMedium] (n1) -- (n5);
\draw[dashed,TUMGrayMedium] (n5) -- (n2);
\draw[dashed,TUMGrayMedium] (n2) -- (n4);
\end{tikzpicture}
    }
    \quad
    \subfloat[Border\label{fig:primalBorderEdges}]
    {
        \begin{tikzpicture}[MyPersp,scale=0.8]
\node[circle,fill,inner sep=0pt, minimum size=2mm,TUMGrayMedium] (nb0) at (0.0,0.0,0.0) {};
\node[circle,fill,inner sep=0pt, minimum size=2mm,TUMGrayMedium] (nb1) at (1.0,0.0,0.0) {};
\node[circle,fill,inner sep=0pt, minimum size=2mm,TUMGrayMedium] (nb2) at (0.0,1.0,0.0) {};
\node[circle,fill,inner sep=0pt, minimum size=2mm,TUMGrayMedium] (nb3) at (1.0,1.0,0.0) {};
\node[circle,fill,inner sep=0pt, minimum size=2mm,TUMGrayMedium] (nb4) at (0.0,0.0,1.0) {};
\node[circle,fill,inner sep=0pt, minimum size=2mm,TUMGrayMedium] (nb5) at (1.0,0.0,1.0) {};
\node[circle,fill,inner sep=0pt, minimum size=2mm,TUMGrayMedium] (nb6) at (0.0,1.0,1.0) {};
\node[circle,fill,inner sep=0pt, minimum size=2mm,TUMGrayMedium] (ni0) at (1.0,1.0,1.0) {};
\node[circle,draw,inner sep=0pt, minimum size=2mm,TUMGrayMedium] (nB0) at (1.5,1.5,1.5) {};
\coordinate (n0) at (1.5,0.0,0.0) ;
\coordinate (n1) at (0.0,1.5,0.0) ;
\coordinate (n2) at (0.0,0.0,1.5) ;
\coordinate (n3) at (1.5,1.5,0.0) ;
\coordinate (n4) at (1.5,0.0,1.5) ;
\coordinate (n5) at (0.0,1.5,1.5) ;
\draw[TUMGrayMedium] (nb0) -- (nb1);
\draw[TUMGrayMedium] (nb1) -- (nb3);
\draw[TUMGrayMedium] (nb3) -- (nb2);
\draw[TUMGrayMedium] (nb2) -- (nb0);
\draw[TUMGrayMedium] (nb4) -- (nb5);
\draw[TUMGrayMedium] (nb5) -- (ni0);
\draw[TUMGrayMedium] (ni0) -- (nb6);
\draw[TUMGrayMedium] (nb6) -- (nb4);
\draw[TUMGrayMedium] (nb0) -- (nb4);
\draw[TUMGrayMedium] (nb1) -- (nb5);
\draw[TUMGrayMedium] (nb2) -- (nb6);
\draw[TUMGrayMedium] (nb3) -- (ni0);
\draw[->-,TUMBlue] (ni0) -- (nB0);
\draw[TUMGrayMedium] (nB0) -- (n4);
\draw[TUMGrayMedium] (n4) -- (nb5);
\draw[TUMGrayMedium] (nB0) -- (n3);
\draw[TUMGrayMedium] (n3) -- (nb3);
\draw[TUMGrayMedium] (nb6) -- (n5);
\draw[TUMGrayMedium] (n5) -- (nB0);
\draw[dashed,TUMGrayMedium] (nb1) -- (n0);
\draw[dashed,TUMGrayMedium] (n0) -- (n3);
\draw[dashed,TUMGrayMedium] (nb2) -- (n1);
\draw[dashed,TUMGrayMedium] (n1) -- (n3);
\draw[dashed,TUMGrayMedium] (nb4) -- (n2);
\draw[dashed,TUMGrayMedium] (n0) -- (n4);
\draw[dashed,TUMGrayMedium] (n1) -- (n5);
\draw[dashed,TUMGrayMedium] (n5) -- (n2);
\draw[dashed,TUMGrayMedium] (n2) -- (n4);
\end{tikzpicture}
    }
    \quad
    \subfloat[Additional border\label{fig:primalAddBorderEdges}]
    {
        \begin{tikzpicture}[MyPersp,scale=0.8]
\node[circle,fill,inner sep=0pt, minimum size=2mm,TUMGrayMedium] (nb0) at (0.0,0.0,0.0) {};
\node[circle,fill,inner sep=0pt, minimum size=2mm,TUMGrayMedium] (nb1) at (1.0,0.0,0.0) {};
\node[circle,fill,inner sep=0pt, minimum size=2mm,TUMGrayMedium] (nb2) at (0.0,1.0,0.0) {};
\node[circle,fill,inner sep=0pt, minimum size=2mm,TUMGrayMedium] (nb3) at (1.0,1.0,0.0) {};
\node[circle,fill,inner sep=0pt, minimum size=2mm,TUMGrayMedium] (nb4) at (0.0,0.0,1.0) {};
\node[circle,fill,inner sep=0pt, minimum size=2mm,TUMGrayMedium] (nb5) at (1.0,0.0,1.0) {};
\node[circle,fill,inner sep=0pt, minimum size=2mm,TUMGrayMedium] (nb6) at (0.0,1.0,1.0) {};
\node[circle,fill,inner sep=0pt, minimum size=2mm,TUMGrayMedium] (ni0) at (1.0,1.0,1.0) {};
\node[circle,draw,inner sep=0pt, minimum size=2mm,TUMGrayMedium] (nB0) at (1.5,1.5,1.5) {};
\coordinate (n0) at (1.5,0.0,0.0) ;
\coordinate (n1) at (0.0,1.5,0.0) ;
\coordinate (n2) at (0.0,0.0,1.5) ;
\coordinate (n3) at (1.5,1.5,0.0) ;
\coordinate (n4) at (1.5,0.0,1.5) ;
\coordinate (n5) at (0.0,1.5,1.5) ;
\draw[TUMGrayMedium] (nb0) -- (nb1);
\draw[TUMGrayMedium] (nb1) -- (nb3);
\draw[TUMGrayMedium] (nb3) -- (nb2);
\draw[TUMGrayMedium] (nb2) -- (nb0);
\draw[TUMGrayMedium] (nb4) -- (nb5);
\draw[TUMGrayMedium] (nb5) -- (ni0);
\draw[TUMGrayMedium] (ni0) -- (nb6);
\draw[TUMGrayMedium] (nb6) -- (nb4);
\draw[TUMGrayMedium] (nb0) -- (nb4);
\draw[TUMGrayMedium] (nb1) -- (nb5);
\draw[TUMGrayMedium] (nb2) -- (nb6);
\draw[TUMGrayMedium] (nb3) -- (ni0);
\draw[TUMGrayMedium] (ni0) -- (nB0);
\draw[->-,TUMBlue] (nB0) -- (n4);
\draw[->-,TUMBlue] (n4) -- (nb5);
\draw[->-,TUMBlue] (nB0) -- (n3);
\draw[->-,TUMBlue] (n3) -- (nb3);
\draw[->-,TUMBlue] (nb6) -- (n5);
\draw[->-,TUMBlue] (n5) -- (nB0);
\draw[dashed,TUMGrayMedium] (nb1) -- (n0);
\draw[dashed,TUMGrayMedium] (n0) -- (n3);
\draw[dashed,TUMGrayMedium] (nb2) -- (n1);
\draw[dashed,TUMGrayMedium] (n1) -- (n3);
\draw[dashed,TUMGrayMedium] (nb4) -- (n2);
\draw[dashed,TUMGrayMedium] (n0) -- (n4);
\draw[dashed,TUMGrayMedium] (n1) -- (n5);
\draw[dashed,TUMGrayMedium] (n5) -- (n2);
\draw[dashed,TUMGrayMedium] (n2) -- (n4);
\end{tikzpicture}
    }
    \caption{Primal edges}
    \label{fig:primalEdges}
\end{figure}

\paragraph*{Inner faces:} 
Faces belonging to windows that are entirely inside the domain, are called inner faces $f_\mathrm{i}\in\mathcal{F}_\mathrm{i}$.

\paragraph*{Border faces:} 
Border faces $f_\mathrm{b}\in\mathcal{F}_\mathrm{b}$ belong to windows that are not completely inside the domain.

\paragraph*{Additional border faces:} 
These faces $f_\mathrm{B}\in\mathcal{F}_\mathrm{B}$ do not belong to a window,
but they are necessery to fill the entire domain with volumes.

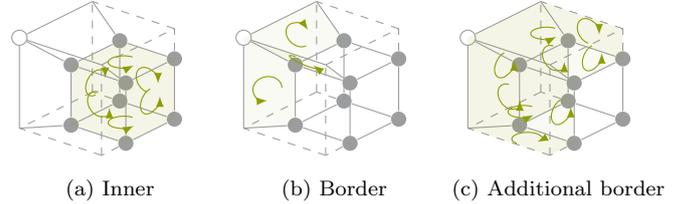
\begin{figure}[!ht]
    \centering
    \subfloat[Inner\label{fig:primalInnerFaces}]
    {
        \begin{tikzpicture}[MyPersp,scale=0.8]
\node[circle,fill,inner sep=0pt, minimum size=2mm,TUMGrayMedium] (nb0) at (0.0,0.0,0.0) {};
\node[circle,fill,inner sep=0pt, minimum size=2mm,TUMGrayMedium] (nb1) at (1.0,0.0,0.0) {};
\node[circle,fill,inner sep=0pt, minimum size=2mm,TUMGrayMedium] (nb2) at (0.0,1.0,0.0) {};
\node[circle,fill,inner sep=0pt, minimum size=2mm,TUMGrayMedium] (nb3) at (1.0,1.0,0.0) {};
\node[circle,fill,inner sep=0pt, minimum size=2mm,TUMGrayMedium] (nb4) at (0.0,0.0,1.0) {};
\node[circle,fill,inner sep=0pt, minimum size=2mm,TUMGrayMedium] (nb5) at (1.0,0.0,1.0) {};
\node[circle,fill,inner sep=0pt, minimum size=2mm,TUMGrayMedium] (nb6) at (0.0,1.0,1.0) {};
\node[circle,fill,inner sep=0pt, minimum size=2mm,TUMGrayMedium] (ni0) at (1.0,1.0,1.0) {};
\node[circle,draw,inner sep=0pt, minimum size=2mm,TUMGrayMedium] (nB0) at (1.5,1.5,1.5) {};
\coordinate (n0) at (1.5,0.0,0.0) ;
\coordinate (n1) at (0.0,1.5,0.0) ;
\coordinate (n2) at (0.0,0.0,1.5) ;
\coordinate (n3) at (1.5,1.5,0.0) ;
\coordinate (n4) at (1.5,0.0,1.5) ;
\coordinate (n5) at (0.0,1.5,1.5) ;
\draw[TUMGrayMedium] (nb0) -- (nb1);
\draw[TUMGrayMedium] (nb1) -- (nb3);
\draw[TUMGrayMedium] (nb3) -- (nb2);
\draw[TUMGrayMedium] (nb2) -- (nb0);
\draw[TUMGrayMedium] (nb4) -- (nb5);
\draw[TUMGrayMedium] (nb5) -- (ni0);
\draw[TUMGrayMedium] (ni0) -- (nb6);
\draw[TUMGrayMedium] (nb6) -- (nb4);
\draw[TUMGrayMedium] (nb0) -- (nb4);
\draw[TUMGrayMedium] (nb1) -- (nb5);
\draw[TUMGrayMedium] (nb2) -- (nb6);
\draw[TUMGrayMedium] (nb3) -- (ni0);
\draw[TUMGrayMedium] (ni0) -- (nB0);
\draw[TUMGrayMedium] (nB0) -- (n4);
\draw[TUMGrayMedium] (n4) -- (nb5);
\draw[TUMGrayMedium] (nB0) -- (n3);
\draw[TUMGrayMedium] (n3) -- (nb3);
\draw[TUMGrayMedium] (nb6) -- (n5);
\draw[TUMGrayMedium] (n5) -- (nB0);
\draw[dashed,TUMGrayMedium] (nb1) -- (n0);
\draw[dashed,TUMGrayMedium] (n0) -- (n3);
\draw[dashed,TUMGrayMedium] (nb2) -- (n1);
\draw[dashed,TUMGrayMedium] (n1) -- (n3);
\draw[dashed,TUMGrayMedium] (nb4) -- (n2);
\draw[dashed,TUMGrayMedium] (n0) -- (n4);
\draw[dashed,TUMGrayMedium] (n1) -- (n5);
\draw[dashed,TUMGrayMedium] (n5) -- (n2);
\draw[dashed,TUMGrayMedium] (n2) -- (n4);
\begin{scope}[canvas is plane={O(0.5,0.5,0.0)x(1.5,0.5,0.0)y(0.5,1.5,0.0)}]
	\node[TUMGreen] (fi0) at (0,0) {};
	\circledarrow{TUMGreen}{fi0}{0.2}
\end{scope}
\fill[TUMGreen,fill opacity=0.05] (nb0.center) -- (nb1.center) -- (nb3.center) -- (nb2.center) -- cycle;
\begin{scope}[canvas is plane={O(1.0,0.5,0.5)x(1.0,1.5,0.5)y(1.0,0.5,1.5)}]
	\node[TUMGreen] (fi1) at (0,0) {};
	\circledarrow{TUMGreen}{fi1}{0.2}
\end{scope}
\fill[TUMGreen,fill opacity=0.05] (nb1.center) -- (nb3.center) -- (ni0.center) -- (nb5.center) -- cycle;
\begin{scope}[canvas is plane={O(0.5,1.0,0.5)x(1.5,1.0,0.5)y(0.5,1.0,-0.5)}]
	\node[TUMGreen] (fi2) at (0,0) {};
	\circledarrow{TUMGreen}{fi2}{0.2}
\end{scope}
\fill[TUMGreen,fill opacity=0.05] (nb3.center) -- (nb2.center) -- (nb6.center) -- (ni0.center) -- cycle;
\begin{scope}[canvas is plane={O(0.0,0.5,0.5)x(0.0,1.5,0.5)y(0.0,0.5,-0.5)}]
	\node[TUMGreen] (fi3) at (0,0) {};
	\circledarrow{TUMGreen}{fi3}{0.2}
\end{scope}
\fill[TUMGreen,fill opacity=0.05] (nb2.center) -- (nb0.center) -- (nb4.center) -- (nb6.center) -- cycle;
\begin{scope}[canvas is plane={O(0.5,0.0,0.5)x(1.5,0.0,0.5)y(0.5,0.0,1.5)}]
	\node[TUMGreen] (fi4) at (0,0) {};
	\circledarrow{TUMGreen}{fi4}{0.2}
\end{scope}
\fill[TUMGreen,fill opacity=0.05] (nb0.center) -- (nb1.center) -- (nb5.center) -- (nb4.center) -- cycle;
\begin{scope}[canvas is plane={O(0.5,0.5,1.0)x(1.5,0.5,1.0)y(0.5,1.5,1.0)}]
	\node[TUMGreen] (fi5) at (0,0) {};
	\circledarrow{TUMGreen}{fi5}{0.2}
\end{scope}
\fill[TUMGreen,fill opacity=0.05] (nb4.center) -- (nb5.center) -- (ni0.center) -- (nb6.center) -- cycle;
\end{tikzpicture}
    }
    \quad
    \subfloat[Border\label{fig:primalBorderFaces}]
    {
        \begin{tikzpicture}[MyPersp,scale=0.8]
\node[circle,fill,inner sep=0pt, minimum size=2mm,TUMGrayMedium] (nb0) at (0.0,0.0,0.0) {};
\node[circle,fill,inner sep=0pt, minimum size=2mm,TUMGrayMedium] (nb1) at (1.0,0.0,0.0) {};
\node[circle,fill,inner sep=0pt, minimum size=2mm,TUMGrayMedium] (nb2) at (0.0,1.0,0.0) {};
\node[circle,fill,inner sep=0pt, minimum size=2mm,TUMGrayMedium] (nb3) at (1.0,1.0,0.0) {};
\node[circle,fill,inner sep=0pt, minimum size=2mm,TUMGrayMedium] (nb4) at (0.0,0.0,1.0) {};
\node[circle,fill,inner sep=0pt, minimum size=2mm,TUMGrayMedium] (nb5) at (1.0,0.0,1.0) {};
\node[circle,fill,inner sep=0pt, minimum size=2mm,TUMGrayMedium] (nb6) at (0.0,1.0,1.0) {};
\node[circle,fill,inner sep=0pt, minimum size=2mm,TUMGrayMedium] (ni0) at (1.0,1.0,1.0) {};
\node[circle,draw,inner sep=0pt, minimum size=2mm,TUMGrayMedium] (nB0) at (1.5,1.5,1.5) {};
\coordinate (n0) at (1.5,0.0,0.0) ;
\coordinate (n1) at (0.0,1.5,0.0) ;
\coordinate (n2) at (0.0,0.0,1.5) ;
\coordinate (n3) at (1.5,1.5,0.0) ;
\coordinate (n4) at (1.5,0.0,1.5) ;
\coordinate (n5) at (0.0,1.5,1.5) ;
\draw[TUMGrayMedium] (nb0) -- (nb1);
\draw[TUMGrayMedium] (nb1) -- (nb3);
\draw[TUMGrayMedium] (nb3) -- (nb2);
\draw[TUMGrayMedium] (nb2) -- (nb0);
\draw[TUMGrayMedium] (nb4) -- (nb5);
\draw[TUMGrayMedium] (nb5) -- (ni0);
\draw[TUMGrayMedium] (ni0) -- (nb6);
\draw[TUMGrayMedium] (nb6) -- (nb4);
\draw[TUMGrayMedium] (nb0) -- (nb4);
\draw[TUMGrayMedium] (nb1) -- (nb5);
\draw[TUMGrayMedium] (nb2) -- (nb6);
\draw[TUMGrayMedium] (nb3) -- (ni0);
\draw[TUMGrayMedium] (ni0) -- (nB0);
\draw[TUMGrayMedium] (nB0) -- (n4);
\draw[TUMGrayMedium] (n4) -- (nb5);
\draw[TUMGrayMedium] (nB0) -- (n3);
\draw[TUMGrayMedium] (n3) -- (nb3);
\draw[TUMGrayMedium] (nb6) -- (n5);
\draw[TUMGrayMedium] (n5) -- (nB0);
\draw[dashed,TUMGrayMedium] (nb1) -- (n0);
\draw[dashed,TUMGrayMedium] (n0) -- (n3);
\draw[dashed,TUMGrayMedium] (nb2) -- (n1);
\draw[dashed,TUMGrayMedium] (n1) -- (n3);
\draw[dashed,TUMGrayMedium] (nb4) -- (n2);
\draw[dashed,TUMGrayMedium] (n0) -- (n4);
\draw[dashed,TUMGrayMedium] (n1) -- (n5);
\draw[dashed,TUMGrayMedium] (n5) -- (n2);
\draw[dashed,TUMGrayMedium] (n2) -- (n4);
\begin{scope}[canvas is plane={O(1.27,0.633,1.27)x(1.27,1.63,1.27)y(1.97,0.633,1.97)}]
	\node[TUMGreen] (fb0) at (0,0) {};
	\circledarrow{TUMGreen}{fb0}{0.2}
\end{scope}
\fill[TUMGreen,fill opacity=0.05] (nb5.center) -- (ni0.center) -- (nB0.center) -- (n4.center) -- cycle;
\begin{scope}[canvas is plane={O(1.27,1.27,0.633)x(1.27,1.27,1.63)y(1.97,1.97,0.633)}]
	\node[TUMGreen] (fb1) at (0,0) {};
	\circledarrow{TUMGreen}{fb1}{0.2}
\end{scope}
\fill[TUMGreen,fill opacity=0.05] (nb3.center) -- (ni0.center) -- (nB0.center) -- (n3.center) -- cycle;
\begin{scope}[canvas is plane={O(0.633,1.27,1.27)x(1.63,1.27,1.27)y(0.633,0.56,0.56)}]
	\node[TUMGreen] (fb2) at (0,0) {};
	\circledarrow{TUMGreen}{fb2}{0.2}
\end{scope}
\fill[TUMGreen,fill opacity=0.05] (ni0.center) -- (nb6.center) -- (n5.center) -- (nB0.center) -- cycle;
\end{tikzpicture}
    }
    \quad
    \subfloat[Additional border\label{fig:primalAddBorderFaces}]
    {
        \begin{tikzpicture}[MyPersp,scale=0.8]
\node[circle,fill,inner sep=0pt, minimum size=2mm,TUMGrayMedium] (nb0) at (0.0,0.0,0.0) {};
\node[circle,fill,inner sep=0pt, minimum size=2mm,TUMGrayMedium] (nb1) at (1.0,0.0,0.0) {};
\node[circle,fill,inner sep=0pt, minimum size=2mm,TUMGrayMedium] (nb2) at (0.0,1.0,0.0) {};
\node[circle,fill,inner sep=0pt, minimum size=2mm,TUMGrayMedium] (nb3) at (1.0,1.0,0.0) {};
\node[circle,fill,inner sep=0pt, minimum size=2mm,TUMGrayMedium] (nb4) at (0.0,0.0,1.0) {};
\node[circle,fill,inner sep=0pt, minimum size=2mm,TUMGrayMedium] (nb5) at (1.0,0.0,1.0) {};
\node[circle,fill,inner sep=0pt, minimum size=2mm,TUMGrayMedium] (nb6) at (0.0,1.0,1.0) {};
\node[circle,fill,inner sep=0pt, minimum size=2mm,TUMGrayMedium] (ni0) at (1.0,1.0,1.0) {};
\node[circle,draw,inner sep=0pt, minimum size=2mm,TUMGrayMedium] (nB0) at (1.5,1.5,1.5) {};
\coordinate (n0) at (1.5,0.0,0.0) ;
\coordinate (n1) at (0.0,1.5,0.0) ;
\coordinate (n2) at (0.0,0.0,1.5) ;
\coordinate (n3) at (1.5,1.5,0.0) ;
\coordinate (n4) at (1.5,0.0,1.5) ;
\coordinate (n5) at (0.0,1.5,1.5) ;
\draw[TUMGrayMedium] (nb0) -- (nb1);
\draw[TUMGrayMedium] (nb1) -- (nb3);
\draw[TUMGrayMedium] (nb3) -- (nb2);
\draw[TUMGrayMedium] (nb2) -- (nb0);
\draw[TUMGrayMedium] (nb4) -- (nb5);
\draw[TUMGrayMedium] (nb5) -- (ni0);
\draw[TUMGrayMedium] (ni0) -- (nb6);
\draw[TUMGrayMedium] (nb6) -- (nb4);
\draw[TUMGrayMedium] (nb0) -- (nb4);
\draw[TUMGrayMedium] (nb1) -- (nb5);
\draw[TUMGrayMedium] (nb2) -- (nb6);
\draw[TUMGrayMedium] (nb3) -- (ni0);
\draw[TUMGrayMedium] (ni0) -- (nB0);
\draw[TUMGrayMedium] (nB0) -- (n4);
\draw[TUMGrayMedium] (n4) -- (nb5);
\draw[TUMGrayMedium] (nB0) -- (n3);
\draw[TUMGrayMedium] (n3) -- (nb3);
\draw[TUMGrayMedium] (nb6) -- (n5);
\draw[TUMGrayMedium] (n5) -- (nB0);
\draw[dashed,TUMGrayMedium] (nb1) -- (n0);
\draw[dashed,TUMGrayMedium] (n0) -- (n3);
\draw[dashed,TUMGrayMedium] (nb2) -- (n1);
\draw[dashed,TUMGrayMedium] (n1) -- (n3);
\draw[dashed,TUMGrayMedium] (nb4) -- (n2);
\draw[dashed,TUMGrayMedium] (n0) -- (n4);
\draw[dashed,TUMGrayMedium] (n1) -- (n5);
\draw[dashed,TUMGrayMedium] (n5) -- (n2);
\draw[dashed,TUMGrayMedium] (n2) -- (n4);
\begin{scope}[canvas is plane={O(1.27,0.633,0.0)x(2.27,0.633,0.0)y(1.27,1.63,0.0)}]
	\node[TUMGreen] (fB0a) at (0,0) {};
	\circledarrow{TUMGreen}{fB0a}{0.2}
\end{scope}
\fill[TUMGreen,fill opacity=0.05] (nb1.center) -- (n0.center) -- (n3.center) -- (nb3.center) -- cycle;
\begin{scope}[canvas is plane={O(1.5,0.75,0.75)x(1.5,1.75,0.75)y(1.5,0.75,-0.25)}]
	\node[TUMGreen] (fB0b) at (0,0) {};
	\circledarrow{TUMGreen}{fB0b}{0.2}
\end{scope}
\fill[TUMGreen,fill opacity=0.05] (n0.center) -- (n4.center) -- (nB0.center) -- (n3.center) -- cycle;
\begin{scope}[canvas is plane={O(1.27,0.0,0.633)x(2.27,0.0,0.633)y(1.27,0.0,-0.367)}]
	\node[TUMGreen] (fB0c) at (0,0) {};
	\circledarrow{TUMGreen}{fB0c}{0.2}
\end{scope}
\fill[TUMGreen,fill opacity=0.05] (nb1.center) -- (nb5.center) -- (n4.center) -- (n0.center) -- cycle;
\begin{scope}[canvas is plane={O(0.633,1.27,0.0)x(1.63,1.27,0.0)y(0.633,0.267,0.0)}]
	\node[TUMGreen] (fB1a) at (0,0) {};
	\circledarrow{TUMGreen}{fB1a}{0.2}
\end{scope}
\fill[TUMGreen,fill opacity=0.05] (nb3.center) -- (nb2.center) -- (n1.center) -- (n3.center) -- cycle;
\begin{scope}[canvas is plane={O(0.75,1.5,0.75)x(1.75,1.5,0.75)y(0.75,1.5,-0.25)}]
	\node[TUMGreen] (fB1b) at (0,0) {};
	\circledarrow{TUMGreen}{fB1b}{0.2}
\end{scope}
\fill[TUMGreen,fill opacity=0.05] (n1.center) -- (n5.center) -- (nB0.center) -- (n3.center) -- cycle;
\begin{scope}[canvas is plane={O(0.0,1.27,0.633)x(0.0,2.27,0.633)y(0.0,1.27,-0.367)}]
	\node[TUMGreen] (fB1c) at (0,0) {};
	\circledarrow{TUMGreen}{fB1c}{0.2}
\end{scope}
\fill[TUMGreen,fill opacity=0.05] (nb2.center) -- (nb6.center) -- (n5.center) -- (n1.center) -- cycle;
\begin{scope}[canvas is plane={O(0.633,0.0,1.27)x(1.63,0.0,1.27)y(0.633,0.0,2.27)}]
	\node[TUMGreen] (fB2a) at (0,0) {};
	\circledarrow{TUMGreen}{fB2a}{0.2}
\end{scope}
\fill[TUMGreen,fill opacity=0.05] (nb4.center) -- (nb5.center) -- (n4.center) -- (n2.center) -- cycle;
\begin{scope}[canvas is plane={O(0.75,0.75,1.5)x(1.75,0.75,1.5)y(0.75,1.75,1.5)}]
	\node[TUMGreen] (fB2b) at (0,0) {};
	\circledarrow{TUMGreen}{fB2b}{0.2}
\end{scope}
\fill[TUMGreen,fill opacity=0.05] (n2.center) -- (n4.center) -- (nB0.center) -- (n5.center) -- cycle;
\begin{scope}[canvas is plane={O(0.0,0.633,1.27)x(0.0,1.63,1.27)y(0.0,0.633,0.267)}]
	\node[TUMGreen] (fB2c) at (0,0) {};
	\circledarrow{TUMGreen}{fB2c}{0.2}
\end{scope}
\fill[TUMGreen,fill opacity=0.05] (nb6.center) -- (nb4.center) -- (n2.center) -- (n5.center) -- cycle;
\end{tikzpicture}
    }
    \caption{Primal faces}
    \label{fig:primalFaces}
\end{figure}

\paragraph*{Inner volumes}
Volumes that lie inside the domain or on the boundary with a \ac{dbc} are inner volumes $v_\mathrm{i}\in\mathcal{V}_\mathrm{i}$.

\paragraph*{Border volumes}
Only volumes at the boundary with a \ac{nbc} are border volumes $v_\mathrm{b}\in\mathcal{V}_\mathrm{b}$.

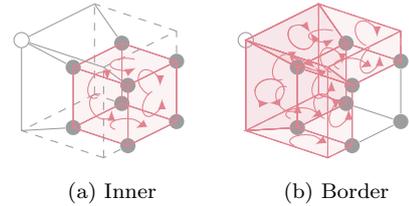
\begin{figure}[!ht]
    \centering
    \subfloat[Inner\label{fig:primalInnerVolumes}]
    {
        \begin{tikzpicture}[MyPersp,scale=0.8]
\node[circle,fill,inner sep=0pt, minimum size=2mm,TUMGrayMedium] (nb0) at (0.0,0.0,0.0) {};
\node[circle,fill,inner sep=0pt, minimum size=2mm,TUMGrayMedium] (nb1) at (1.0,0.0,0.0) {};
\node[circle,fill,inner sep=0pt, minimum size=2mm,TUMGrayMedium] (nb2) at (0.0,1.0,0.0) {};
\node[circle,fill,inner sep=0pt, minimum size=2mm,TUMGrayMedium] (nb3) at (1.0,1.0,0.0) {};
\node[circle,fill,inner sep=0pt, minimum size=2mm,TUMGrayMedium] (nb4) at (0.0,0.0,1.0) {};
\node[circle,fill,inner sep=0pt, minimum size=2mm,TUMGrayMedium] (nb5) at (1.0,0.0,1.0) {};
\node[circle,fill,inner sep=0pt, minimum size=2mm,TUMGrayMedium] (nb6) at (0.0,1.0,1.0) {};
\node[circle,fill,inner sep=0pt, minimum size=2mm,TUMGrayMedium] (ni0) at (1.0,1.0,1.0) {};
\node[circle,draw,inner sep=0pt, minimum size=2mm,TUMGrayMedium] (nB0) at (1.5,1.5,1.5) {};
\coordinate (n0) at (1.5,0.0,0.0) ;
\coordinate (n1) at (0.0,1.5,0.0) ;
\coordinate (n2) at (0.0,0.0,1.5) ;
\coordinate (n3) at (1.5,1.5,0.0) ;
\coordinate (n4) at (1.5,0.0,1.5) ;
\coordinate (n5) at (0.0,1.5,1.5) ;
\draw[TUMGrayMedium] (nb0) -- (nb1);
\draw[TUMGrayMedium] (nb1) -- (nb3);
\draw[TUMGrayMedium] (nb3) -- (nb2);
\draw[TUMGrayMedium] (nb2) -- (nb0);
\draw[TUMGrayMedium] (nb4) -- (nb5);
\draw[TUMGrayMedium] (nb5) -- (ni0);
\draw[TUMGrayMedium] (ni0) -- (nb6);
\draw[TUMGrayMedium] (nb6) -- (nb4);
\draw[TUMGrayMedium] (nb0) -- (nb4);
\draw[TUMGrayMedium] (nb1) -- (nb5);
\draw[TUMGrayMedium] (nb2) -- (nb6);
\draw[TUMGrayMedium] (nb3) -- (ni0);
\draw[TUMGrayMedium] (ni0) -- (nB0);
\draw[TUMGrayMedium] (nB0) -- (n4);
\draw[TUMGrayMedium] (n4) -- (nb5);
\draw[TUMGrayMedium] (nB0) -- (n3);
\draw[TUMGrayMedium] (n3) -- (nb3);
\draw[TUMGrayMedium] (nb6) -- (n5);
\draw[TUMGrayMedium] (n5) -- (nB0);
\draw[dashed,TUMGrayMedium] (nb1) -- (n0);
\draw[dashed,TUMGrayMedium] (n0) -- (n3);
\draw[dashed,TUMGrayMedium] (nb2) -- (n1);
\draw[dashed,TUMGrayMedium] (n1) -- (n3);
\draw[dashed,TUMGrayMedium] (nb4) -- (n2);
\draw[dashed,TUMGrayMedium] (n0) -- (n4);
\draw[dashed,TUMGrayMedium] (n1) -- (n5);
\draw[dashed,TUMGrayMedium] (n5) -- (n2);
\draw[dashed,TUMGrayMedium] (n2) -- (n4);
\begin{scope}[canvas is plane={O(0.5,0.5,0.0)x(1.5,0.5,0.0)y(0.5,-0.5,0.0)}]
	\node[TUMRose] (fi0) at (0,0) {};
	\circledarrow{TUMRose}{fi0}{0.2}
\end{scope}
\filldraw[TUMRose,fill opacity=0.05] (nb2.center) -- (nb3.center) -- (nb1.center) -- (nb0.center) -- cycle;
\begin{scope}[canvas is plane={O(1.0,0.5,0.5)x(1.0,1.5,0.5)y(1.0,0.5,1.5)}]
	\node[TUMRose] (fi1) at (0,0) {};
	\circledarrow{TUMRose}{fi1}{0.2}
\end{scope}
\filldraw[TUMRose,fill opacity=0.05] (nb1.center) -- (nb3.center) -- (ni0.center) -- (nb5.center) -- cycle;
\begin{scope}[canvas is plane={O(0.5,1.0,0.5)x(1.5,1.0,0.5)y(0.5,1.0,-0.5)}]
	\node[TUMRose] (fi2) at (0,0) {};
	\circledarrow{TUMRose}{fi2}{0.2}
\end{scope}
\filldraw[TUMRose,fill opacity=0.05] (nb3.center) -- (nb2.center) -- (nb6.center) -- (ni0.center) -- cycle;
\begin{scope}[canvas is plane={O(0.0,0.5,0.5)x(0.0,1.5,0.5)y(0.0,0.5,-0.5)}]
	\node[TUMRose] (fi3) at (0,0) {};
	\circledarrow{TUMRose}{fi3}{0.2}
\end{scope}
\filldraw[TUMRose,fill opacity=0.05] (nb2.center) -- (nb0.center) -- (nb4.center) -- (nb6.center) -- cycle;
\begin{scope}[canvas is plane={O(0.5,0.0,0.5)x(1.5,0.0,0.5)y(0.5,0.0,1.5)}]
	\node[TUMRose] (fi4) at (0,0) {};
	\circledarrow{TUMRose}{fi4}{0.2}
\end{scope}
\filldraw[TUMRose,fill opacity=0.05] (nb0.center) -- (nb1.center) -- (nb5.center) -- (nb4.center) -- cycle;
\begin{scope}[canvas is plane={O(0.5,0.5,1.0)x(1.5,0.5,1.0)y(0.5,1.5,1.0)}]
	\node[TUMRose] (fi5) at (0,0) {};
	\circledarrow{TUMRose}{fi5}{0.2}
\end{scope}
\filldraw[TUMRose,fill opacity=0.05] (nb4.center) -- (nb5.center) -- (ni0.center) -- (nb6.center) -- cycle;
\end{tikzpicture}
    }
    \quad
    \subfloat[Border\label{fig:primalBorderVolumes}]
    {
        \begin{tikzpicture}[MyPersp,scale=0.8]
\node[circle,fill,inner sep=0pt, minimum size=2mm,TUMGrayMedium] (nb0) at (0.0,0.0,0.0) {};
\node[circle,fill,inner sep=0pt, minimum size=2mm,TUMGrayMedium] (nb1) at (1.0,0.0,0.0) {};
\node[circle,fill,inner sep=0pt, minimum size=2mm,TUMGrayMedium] (nb2) at (0.0,1.0,0.0) {};
\node[circle,fill,inner sep=0pt, minimum size=2mm,TUMGrayMedium] (nb3) at (1.0,1.0,0.0) {};
\node[circle,fill,inner sep=0pt, minimum size=2mm,TUMGrayMedium] (nb4) at (0.0,0.0,1.0) {};
\node[circle,fill,inner sep=0pt, minimum size=2mm,TUMGrayMedium] (nb5) at (1.0,0.0,1.0) {};
\node[circle,fill,inner sep=0pt, minimum size=2mm,TUMGrayMedium] (nb6) at (0.0,1.0,1.0) {};
\node[circle,fill,inner sep=0pt, minimum size=2mm,TUMGrayMedium] (ni0) at (1.0,1.0,1.0) {};
\node[circle,draw,inner sep=0pt, minimum size=2mm,TUMGrayMedium] (nB0) at (1.5,1.5,1.5) {};
\coordinate (n0) at (1.5,0.0,0.0) ;
\coordinate (n1) at (0.0,1.5,0.0) ;
\coordinate (n2) at (0.0,0.0,1.5) ;
\coordinate (n3) at (1.5,1.5,0.0) ;
\coordinate (n4) at (1.5,0.0,1.5) ;
\coordinate (n5) at (0.0,1.5,1.5) ;
\draw[TUMGrayMedium] (nb0) -- (nb1);
\draw[TUMGrayMedium] (nb1) -- (nb3);
\draw[TUMGrayMedium] (nb3) -- (nb2);
\draw[TUMGrayMedium] (nb2) -- (nb0);
\draw[TUMGrayMedium] (nb4) -- (nb5);
\draw[TUMGrayMedium] (nb5) -- (ni0);
\draw[TUMGrayMedium] (ni0) -- (nb6);
\draw[TUMGrayMedium] (nb6) -- (nb4);
\draw[TUMGrayMedium] (nb0) -- (nb4);
\draw[TUMGrayMedium] (nb1) -- (nb5);
\draw[TUMGrayMedium] (nb2) -- (nb6);
\draw[TUMGrayMedium] (nb3) -- (ni0);
\draw[TUMGrayMedium] (ni0) -- (nB0);
\draw[TUMGrayMedium] (nB0) -- (n4);
\draw[TUMGrayMedium] (n4) -- (nb5);
\draw[TUMGrayMedium] (nB0) -- (n3);
\draw[TUMGrayMedium] (n3) -- (nb3);
\draw[TUMGrayMedium] (nb6) -- (n5);
\draw[TUMGrayMedium] (n5) -- (nB0);
\draw[dashed,TUMGrayMedium] (nb1) -- (n0);
\draw[dashed,TUMGrayMedium] (n0) -- (n3);
\draw[dashed,TUMGrayMedium] (nb2) -- (n1);
\draw[dashed,TUMGrayMedium] (n1) -- (n3);
\draw[dashed,TUMGrayMedium] (nb4) -- (n2);
\draw[dashed,TUMGrayMedium] (n0) -- (n4);
\draw[dashed,TUMGrayMedium] (n1) -- (n5);
\draw[dashed,TUMGrayMedium] (n5) -- (n2);
\draw[dashed,TUMGrayMedium] (n2) -- (n4);
\begin{scope}[canvas is plane={O(1.0,0.5,0.5)x(1.0,1.5,0.5)y(1.0,0.5,-0.5)}]
	\node[TUMRose] (fi1) at (0,0) {};
	\circledarrow{TUMRose}{fi1}{0.2}
\end{scope}
\filldraw[TUMRose,fill opacity=0.05] (nb5.center) -- (ni0.center) -- (nb3.center) -- (nb1.center) -- cycle;
\begin{scope}[canvas is plane={O(1.27,0.633,0.0)x(2.27,0.633,0.0)y(1.27,-0.367,0.0)}]
	\node[TUMRose] (fB0a) at (0,0) {};
	\circledarrow{TUMRose}{fB0a}{0.2}
\end{scope}
\filldraw[TUMRose,fill opacity=0.05] (nb3.center) -- (n3.center) -- (n0.center) -- (nb1.center) -- cycle;
\begin{scope}[canvas is plane={O(1.5,0.75,0.75)x(1.5,1.75,0.75)y(1.5,0.75,1.75)}]
	\node[TUMRose] (fB0b) at (0,0) {};
	\circledarrow{TUMRose}{fB0b}{0.2}
\end{scope}
\filldraw[TUMRose,fill opacity=0.05] (n3.center) -- (nB0.center) -- (n4.center) -- (n0.center) -- cycle;
\begin{scope}[canvas is plane={O(1.27,0.0,0.633)x(2.27,0.0,0.633)y(1.27,0.0,1.63)}]
	\node[TUMRose] (fB0c) at (0,0) {};
	\circledarrow{TUMRose}{fB0c}{0.2}
\end{scope}
\filldraw[TUMRose,fill opacity=0.05] (n0.center) -- (n4.center) -- (nb5.center) -- (nb1.center) -- cycle;
\begin{scope}[canvas is plane={O(1.27,0.633,1.27)x(1.27,1.63,1.27)y(0.56,0.633,0.56)}]
	\node[TUMRose] (fb0) at (0,0) {};
	\circledarrow{TUMRose}{fb0}{0.2}
\end{scope}
\filldraw[TUMRose,fill opacity=0.05] (n4.center) -- (nB0.center) -- (ni0.center) -- (nb5.center) -- cycle;
\begin{scope}[canvas is plane={O(1.27,1.27,0.633)x(1.27,1.27,1.63)y(1.97,1.97,0.633)}]
	\node[TUMRose] (fb1) at (0,0) {};
	\circledarrow{TUMRose}{fb1}{0.2}
\end{scope}
\filldraw[TUMRose,fill opacity=0.05] (nb3.center) -- (ni0.center) -- (nB0.center) -- (n3.center) -- cycle;
\begin{scope}[canvas is plane={O(1.27,1.27,0.633)x(1.27,1.27,1.63)y(0.56,0.56,0.633)}]
	\node[TUMRose] (fb1) at (0,0) {};
	\circledarrow{TUMRose}{fb1}{0.2}
\end{scope}
\filldraw[TUMRose,fill opacity=0.05] (n3.center) -- (nB0.center) -- (ni0.center) -- (nb3.center) -- cycle;
\begin{scope}[canvas is plane={O(0.633,1.27,0.0)x(1.63,1.27,0.0)y(0.633,0.267,0.0)}]
	\node[TUMRose] (fB1a) at (0,0) {};
	\circledarrow{TUMRose}{fB1a}{0.2}
\end{scope}
\filldraw[TUMRose,fill opacity=0.05] (nb3.center) -- (nb2.center) -- (n1.center) -- (n3.center) -- cycle;
\begin{scope}[canvas is plane={O(0.75,1.5,0.75)x(1.75,1.5,0.75)y(0.75,1.5,-0.25)}]
	\node[TUMRose] (fB1b) at (0,0) {};
	\circledarrow{TUMRose}{fB1b}{0.2}
\end{scope}
\filldraw[TUMRose,fill opacity=0.05] (n1.center) -- (n5.center) -- (nB0.center) -- (n3.center) -- cycle;
\begin{scope}[canvas is plane={O(0.0,1.27,0.633)x(0.0,2.27,0.633)y(0.0,1.27,-0.367)}]
	\node[TUMRose] (fB1c) at (0,0) {};
	\circledarrow{TUMRose}{fB1c}{0.2}
\end{scope}
\filldraw[TUMRose,fill opacity=0.05] (nb2.center) -- (nb6.center) -- (n5.center) -- (n1.center) -- cycle;
\begin{scope}[canvas is plane={O(0.633,1.27,1.27)x(1.63,1.27,1.27)y(0.633,1.97,1.97)}]
	\node[TUMRose] (fb2) at (0,0) {};
	\circledarrow{TUMRose}{fb2}{0.2}
\end{scope}
\filldraw[TUMRose,fill opacity=0.05] (nB0.center) -- (n5.center) -- (nb6.center) -- (ni0.center) -- cycle;
\begin{scope}[canvas is plane={O(0.5,1.0,0.5)x(1.5,1.0,0.5)y(0.5,1.0,1.5)}]
	\node[TUMRose] (fi2) at (0,0) {};
	\circledarrow{TUMRose}{fi2}{0.2}
\end{scope}
\filldraw[TUMRose,fill opacity=0.05] (ni0.center) -- (nb6.center) -- (nb2.center) -- (nb3.center) -- cycle;
\begin{scope}[canvas is plane={O(0.633,0.0,1.27)x(1.63,0.0,1.27)y(0.633,0.0,2.27)}]
	\node[TUMRose] (fB2a) at (0,0) {};
	\circledarrow{TUMRose}{fB2a}{0.2}
\end{scope}
\filldraw[TUMRose,fill opacity=0.05] (nb4.center) -- (nb5.center) -- (n4.center) -- (n2.center) -- cycle;
\begin{scope}[canvas is plane={O(0.75,0.75,1.5)x(1.75,0.75,1.5)y(0.75,1.75,1.5)}]
	\node[TUMRose] (fB2b) at (0,0) {};
	\circledarrow{TUMRose}{fB2b}{0.2}
\end{scope}
\filldraw[TUMRose,fill opacity=0.05] (n2.center) -- (n4.center) -- (nB0.center) -- (n5.center) -- cycle;
\begin{scope}[canvas is plane={O(0.0,0.633,1.27)x(0.0,1.63,1.27)y(0.0,0.633,0.267)}]
	\node[TUMRose] (fB2c) at (0,0) {};
	\circledarrow{TUMRose}{fB2c}{0.2}
\end{scope}
\filldraw[TUMRose,fill opacity=0.05] (nb6.center) -- (nb4.center) -- (n2.center) -- (n5.center) -- cycle;
\begin{scope}[canvas is plane={O(0.5,0.5,1.0)x(1.5,0.5,1.0)y(0.5,-0.5,1.0)}]
	\node[TUMRose] (fi5) at (0,0) {};
	\circledarrow{TUMRose}{fi5}{0.2}
\end{scope}
\filldraw[TUMRose,fill opacity=0.05] (nb6.center) -- (ni0.center) -- (nb5.center) -- (nb4.center) -- cycle;
\begin{scope}[canvas is plane={O(0.633,1.27,1.27)x(1.63,1.27,1.27)y(0.633,0.56,0.56)}]
	\node[TUMRose] (fb2) at (0,0) {};
	\circledarrow{TUMRose}{fb2}{0.2}
\end{scope}
\filldraw[TUMRose,fill opacity=0.05] (ni0.center) -- (nb6.center) -- (n5.center) -- (nB0.center) -- cycle;
\begin{scope}[canvas is plane={O(1.27,0.633,1.27)x(1.27,1.63,1.27)y(1.97,0.633,1.97)}]
	\node[TUMRose] (fb0) at (0,0) {};
	\circledarrow{TUMRose}{fb0}{0.2}
\end{scope}
\filldraw[TUMRose,fill opacity=0.05] (nb5.center) -- (ni0.center) -- (nB0.center) -- (n4.center) -- cycle;
\end{tikzpicture}
    }
    \caption{Primal volumes}
    \label{fig:primalVolumes}
\end{figure}

\begin{remark}
On first sight it may seem, that some nodes in Fig. \ref{fig:primalComplex}, 
especially at the corners, are missing.
However, they were left out intentionally. 
Similar to the additional boundary edges in the 2D case in \cite[Fig. 5]{Kotyczka2017Dis}, 
that have no nodes at the corner of the face,
the volumes in the 3D case can also have corners without nodes.
In 3D, there can even be kinks in the faces without having a ``real'' edge at that position.
These kinks are drawn with dotted lines and lie \emph{inside} a face
and have therefore no effect on the result of the \emph{boundary} operator applied to the face.
\end{remark}

\begin{remark}
The categorization differs from \cite{Kotyczka2017Dis}, 
because the physical variables are assigned to the geometric objects in another way.
This is because the energy balance is evaluated on the dual volumes instead of the primal faces.
Subsequently, the driving force is evaluated on the primal instead of the dual edges.
\end{remark}

\subsection{Construction of the Dual 3-Complex\label{subsec:dual-complex}}
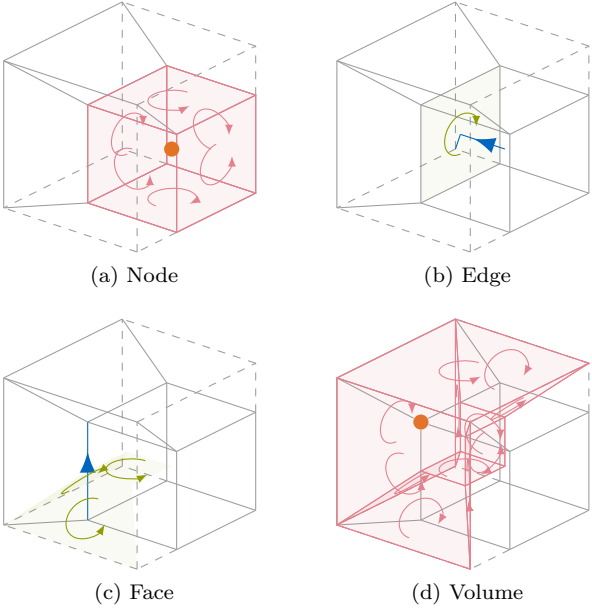
\begin{figure}[!ht]
    \centering
    \subfloat[Node\label{fig:dualNode}]
    {
        \begin{tikzpicture}[MyPersp,scale=1.3]
\coordinate (nb0) at (0.0,0.0,0.0) ;
\coordinate (nb1) at (1.0,0.0,0.0) ;
\coordinate (nb2) at (0.0,1.0,0.0) ;
\coordinate (nb3) at (1.0,1.0,0.0) ;
\coordinate (nb4) at (0.0,0.0,1.0) ;
\coordinate (nb5) at (1.0,0.0,1.0) ;
\coordinate (nb6) at (0.0,1.0,1.0) ;
\coordinate (ni0) at (1.0,1.0,1.0) ;
\coordinate (nB0) at (1.5,1.5,1.5) ;
\coordinate (n0) at (1.5,0.0,0.0) ;
\coordinate (n1) at (0.0,1.5,0.0) ;
\coordinate (n2) at (0.0,0.0,1.5) ;
\coordinate (n3) at (1.5,1.5,0.0) ;
\coordinate (n4) at (1.5,0.0,1.5) ;
\coordinate (n5) at (0.0,1.5,1.5) ;
\draw[TUMGrayMedium] (nb0) -- (nb1);
\draw[TUMGrayMedium] (nb1) -- (nb3);
\draw[TUMGrayMedium] (nb3) -- (nb2);
\draw[TUMGrayMedium] (nb2) -- (nb0);
\draw[TUMGrayMedium] (nb4) -- (nb5);
\draw[TUMGrayMedium] (nb5) -- (ni0);
\draw[TUMGrayMedium] (ni0) -- (nb6);
\draw[TUMGrayMedium] (nb6) -- (nb4);
\draw[TUMGrayMedium] (nb0) -- (nb4);
\draw[TUMGrayMedium] (nb1) -- (nb5);
\draw[TUMGrayMedium] (nb2) -- (nb6);
\draw[TUMGrayMedium] (nb3) -- (ni0);
\draw[TUMGrayMedium] (ni0) -- (nB0);
\draw[TUMGrayMedium] (nB0) -- (n4);
\draw[TUMGrayMedium] (n4) -- (nb5);
\draw[TUMGrayMedium] (nB0) -- (n3);
\draw[TUMGrayMedium] (n3) -- (nb3);
\draw[TUMGrayMedium] (nb6) -- (n5);
\draw[TUMGrayMedium] (n5) -- (nB0);
\draw[dashed,TUMGrayMedium] (nb1) -- (n0);
\draw[dashed,TUMGrayMedium] (n0) -- (n3);
\draw[dashed,TUMGrayMedium] (nb2) -- (n1);
\draw[dashed,TUMGrayMedium] (n1) -- (n3);
\draw[dashed,TUMGrayMedium] (nb4) -- (n2);
\draw[dashed,TUMGrayMedium] (n0) -- (n4);
\draw[dashed,TUMGrayMedium] (n1) -- (n5);
\draw[dashed,TUMGrayMedium] (n5) -- (n2);
\draw[dashed,TUMGrayMedium] (n2) -- (n4);
\begin{scope}[canvas is plane={O(0.5,0.5,0.0)x(1.5,0.5,0.0)y(0.5,-0.5,0.0)}]
	\node[TUMRose] (fi0) at (0,0) {};
	\circledarrow{TUMRose}{fi0}{0.2}
\end{scope}
\filldraw[TUMRose,fill opacity=0.05] (nb2.center) -- (nb3.center) -- (nb1.center) -- (nb0.center) -- cycle;
\begin{scope}[canvas is plane={O(1.0,0.5,0.5)x(1.0,1.5,0.5)y(1.0,0.5,1.5)}]
	\node[TUMRose] (fi1) at (0,0) {};
	\circledarrow{TUMRose}{fi1}{0.2}
\end{scope}
\filldraw[TUMRose,fill opacity=0.05] (nb1.center) -- (nb3.center) -- (ni0.center) -- (nb5.center) -- cycle;
\begin{scope}[canvas is plane={O(0.5,1.0,0.5)x(1.5,1.0,0.5)y(0.5,1.0,-0.5)}]
	\node[TUMRose] (fi2) at (0,0) {};
	\circledarrow{TUMRose}{fi2}{0.2}
\end{scope}
\filldraw[TUMRose,fill opacity=0.05] (nb3.center) -- (nb2.center) -- (nb6.center) -- (ni0.center) -- cycle;
\begin{scope}[canvas is plane={O(0.0,0.5,0.5)x(0.0,1.5,0.5)y(0.0,0.5,-0.5)}]
	\node[TUMRose] (fi3) at (0,0) {};
	\circledarrow{TUMRose}{fi3}{0.2}
\end{scope}
\filldraw[TUMRose,fill opacity=0.05] (nb2.center) -- (nb0.center) -- (nb4.center) -- (nb6.center) -- cycle;
\begin{scope}[canvas is plane={O(0.5,0.0,0.5)x(1.5,0.0,0.5)y(0.5,0.0,1.5)}]
	\node[TUMRose] (fi4) at (0,0) {};
	\circledarrow{TUMRose}{fi4}{0.2}
\end{scope}
\filldraw[TUMRose,fill opacity=0.05] (nb0.center) -- (nb1.center) -- (nb5.center) -- (nb4.center) -- cycle;
\begin{scope}[canvas is plane={O(0.5,0.5,1.0)x(1.5,0.5,1.0)y(0.5,1.5,1.0)}]
	\node[TUMRose] (fi5) at (0,0) {};
	\circledarrow{TUMRose}{fi5}{0.2}
\end{scope}
\filldraw[TUMRose,fill opacity=0.05] (nb4.center) -- (nb5.center) -- (ni0.center) -- (nb6.center) -- cycle;
\node[circle,fill,inner sep=0pt, minimum size=2mm,TUMOrange] (ni0) at (0.5,0.5,0.5) {};
\coordinate (nb0) at (1.5,0.0,0.0) ;
\coordinate (nb1) at (0.0,1.5,0.0) ;
\coordinate (nb2) at (0.0,0.0,1.5) ;
\coordinate (nB0) at (0.5,0.5,0.0) ;
\coordinate (nB3) at (0.0,0.5,0.5) ;
\coordinate (nB4) at (0.5,0.0,0.5) ;
\coordinate (n0) at (1.0,0.5,1.0) ;
\coordinate (n1) at (1.5,0.0,1.5) ;
\coordinate (n2) at (1.0,0.5,0.5) ;
\coordinate (n3) at (0.5,0.5,1.0) ;
\coordinate (n4) at (0.5,1.0,1.0) ;
\coordinate (n5) at (0.0,1.5,1.5) ;
\coordinate (n6) at (0.5,1.0,0.5) ;
\coordinate (n7) at (1.0,1.0,0.5) ;
\coordinate (n8) at (1.5,1.5,0.0) ;
\coordinate (n9) at (1.5,1.5,1.5) ;
\coordinate (n10) at (0.5,0.0,0.0) ;
\coordinate (n11) at (1.0,0.5,0.0) ;
\coordinate (n12) at (0.5,1.0,0.0) ;
\coordinate (n13) at (0.0,0.5,0.0) ;
\coordinate (n14) at (0.5,0.0,1.0) ;
\coordinate (n15) at (0.0,0.5,1.0) ;
\coordinate (n16) at (0.0,0.0,0.5) ;
\coordinate (n17) at (1.0,0.0,0.5) ;
\coordinate (n18) at (0.0,1.0,0.5) ;
\coordinate (n19) at (0.0,0.0,0.0) ;
\coordinate (n20) at (1.0,0.0,0.0) ;
\coordinate (n21) at (0.0,1.0,0.0) ;
\coordinate (n22) at (0.0,0.0,1.0) ;
\end{tikzpicture}
    }
    \qquad
    \subfloat[Edge\label{fig:dualEdge}]
    {
        \begin{tikzpicture}[MyPersp,scale=1.3]
\coordinate (nb0) at (0.0,0.0,0.0) ;
\coordinate (nb1) at (1.0,0.0,0.0) ;
\coordinate (nb2) at (0.0,1.0,0.0) ;
\coordinate (nb3) at (1.0,1.0,0.0) ;
\coordinate (nb4) at (0.0,0.0,1.0) ;
\coordinate (nb5) at (1.0,0.0,1.0) ;
\coordinate (nb6) at (0.0,1.0,1.0) ;
\coordinate (ni0) at (1.0,1.0,1.0) ;
\coordinate (nB0) at (1.5,1.5,1.5) ;
\coordinate (n0) at (1.5,0.0,0.0) ;
\coordinate (n1) at (0.0,1.5,0.0) ;
\coordinate (n2) at (0.0,0.0,1.5) ;
\coordinate (n3) at (1.5,1.5,0.0) ;
\coordinate (n4) at (1.5,0.0,1.5) ;
\coordinate (n5) at (0.0,1.5,1.5) ;
\draw[TUMGrayMedium] (nb0) -- (nb1);
\draw[TUMGrayMedium] (nb1) -- (nb3);
\draw[TUMGrayMedium] (nb3) -- (nb2);
\draw[TUMGrayMedium] (nb2) -- (nb0);
\draw[TUMGrayMedium] (nb4) -- (nb5);
\draw[TUMGrayMedium] (nb5) -- (ni0);
\draw[TUMGrayMedium] (ni0) -- (nb6);
\draw[TUMGrayMedium] (nb6) -- (nb4);
\draw[TUMGrayMedium] (nb0) -- (nb4);
\draw[TUMGrayMedium] (nb1) -- (nb5);
\draw[TUMGrayMedium] (nb2) -- (nb6);
\draw[TUMGrayMedium] (nb3) -- (ni0);
\draw[TUMGrayMedium] (ni0) -- (nB0);
\draw[TUMGrayMedium] (nB0) -- (n4);
\draw[TUMGrayMedium] (n4) -- (nb5);
\draw[TUMGrayMedium] (nB0) -- (n3);
\draw[TUMGrayMedium] (n3) -- (nb3);
\draw[TUMGrayMedium] (nb6) -- (n5);
\draw[TUMGrayMedium] (n5) -- (nB0);
\draw[dashed,TUMGrayMedium] (nb1) -- (n0);
\draw[dashed,TUMGrayMedium] (n0) -- (n3);
\draw[dashed,TUMGrayMedium] (nb2) -- (n1);
\draw[dashed,TUMGrayMedium] (n1) -- (n3);
\draw[dashed,TUMGrayMedium] (nb4) -- (n2);
\draw[dashed,TUMGrayMedium] (n0) -- (n4);
\draw[dashed,TUMGrayMedium] (n1) -- (n5);
\draw[dashed,TUMGrayMedium] (n5) -- (n2);
\draw[dashed,TUMGrayMedium] (n2) -- (n4);
\begin{scope}[canvas is plane={O(1.0,0.5,0.5)x(1.0,1.5,0.5)y(1.0,0.5,1.5)}]
	\node[TUMGreen] (fi1) at (0,0) {};
	\circledarrow{TUMGreen}{fi1}{0.2}
\end{scope}
\fill[TUMGreen,fill opacity=0.05] (nb1.center) -- (nb3.center) -- (ni0.center) -- (nb5.center) -- cycle;
\coordinate (ni0) at (0.5,0.5,0.5) ;
\coordinate (nb0) at (1.5,0.0,0.0) ;
\coordinate (nb1) at (0.0,1.5,0.0) ;
\coordinate (nb2) at (0.0,0.0,1.5) ;
\coordinate (nB0) at (0.5,0.5,0.0) ;
\coordinate (nB3) at (0.0,0.5,0.5) ;
\coordinate (nB4) at (0.5,0.0,0.5) ;
\coordinate (n0) at (1.0,0.5,1.0) ;
\coordinate (n1) at (1.5,0.0,1.5) ;
\coordinate (n2) at (1.0,0.5,0.5) ;
\coordinate (n3) at (0.5,0.5,1.0) ;
\coordinate (n4) at (0.5,1.0,1.0) ;
\coordinate (n5) at (0.0,1.5,1.5) ;
\coordinate (n6) at (0.5,1.0,0.5) ;
\coordinate (n7) at (1.0,1.0,0.5) ;
\coordinate (n8) at (1.5,1.5,0.0) ;
\coordinate (n9) at (1.5,1.5,1.5) ;
\coordinate (n10) at (0.5,0.0,0.0) ;
\coordinate (n11) at (1.0,0.5,0.0) ;
\coordinate (n12) at (0.5,1.0,0.0) ;
\coordinate (n13) at (0.0,0.5,0.0) ;
\coordinate (n14) at (0.5,0.0,1.0) ;
\coordinate (n15) at (0.0,0.5,1.0) ;
\coordinate (n16) at (0.0,0.0,0.5) ;
\coordinate (n17) at (1.0,0.0,0.5) ;
\coordinate (n18) at (0.0,1.0,0.5) ;
\coordinate (n19) at (0.0,0.0,0.0) ;
\coordinate (n20) at (1.0,0.0,0.0) ;
\coordinate (n21) at (0.0,1.0,0.0) ;
\coordinate (n22) at (0.0,0.0,1.0) ;
\draw[->-,TUMBlue] (ni0) -- (n2);
\draw[TUMBlue] (n2) -- (nb0);
\end{tikzpicture}
    }
    \\
    \subfloat[Face\label{fig:dualFace}]
    {
        \begin{tikzpicture}[MyPersp,scale=1.3]
\coordinate (nb0) at (0.0,0.0,0.0) ;
\coordinate (nb1) at (1.0,0.0,0.0) ;
\coordinate (nb2) at (0.0,1.0,0.0) ;
\coordinate (nb3) at (1.0,1.0,0.0) ;
\coordinate (nb4) at (0.0,0.0,1.0) ;
\coordinate (nb5) at (1.0,0.0,1.0) ;
\coordinate (nb6) at (0.0,1.0,1.0) ;
\coordinate (ni0) at (1.0,1.0,1.0) ;
\coordinate (nB0) at (1.5,1.5,1.5) ;
\coordinate (n0) at (1.5,0.0,0.0) ;
\coordinate (n1) at (0.0,1.5,0.0) ;
\coordinate (n2) at (0.0,0.0,1.5) ;
\coordinate (n3) at (1.5,1.5,0.0) ;
\coordinate (n4) at (1.5,0.0,1.5) ;
\coordinate (n5) at (0.0,1.5,1.5) ;
\draw[TUMGrayMedium] (nb0) -- (nb1);
\draw[TUMGrayMedium] (nb1) -- (nb3);
\draw[TUMGrayMedium] (nb3) -- (nb2);
\draw[TUMGrayMedium] (nb2) -- (nb0);
\draw[TUMGrayMedium] (nb4) -- (nb5);
\draw[TUMGrayMedium] (nb5) -- (ni0);
\draw[TUMGrayMedium] (ni0) -- (nb6);
\draw[TUMGrayMedium] (nb6) -- (nb4);
\draw[TUMGrayMedium] (nb0) -- (nb4);
\draw[TUMGrayMedium] (nb1) -- (nb5);
\draw[TUMGrayMedium] (nb2) -- (nb6);
\draw[->-,TUMBlue] (nb3) -- (ni0);
\draw[TUMGrayMedium] (ni0) -- (nB0);
\draw[TUMGrayMedium] (nB0) -- (n4);
\draw[TUMGrayMedium] (n4) -- (nb5);
\draw[TUMGrayMedium] (nB0) -- (n3);
\draw[TUMGrayMedium] (n3) -- (nb3);
\draw[TUMGrayMedium] (nb6) -- (n5);
\draw[TUMGrayMedium] (n5) -- (nB0);
\draw[dashed,TUMGrayMedium] (nb1) -- (n0);
\draw[dashed,TUMGrayMedium] (n0) -- (n3);
\draw[dashed,TUMGrayMedium] (nb2) -- (n1);
\draw[dashed,TUMGrayMedium] (n1) -- (n3);
\draw[dashed,TUMGrayMedium] (nb4) -- (n2);
\draw[dashed,TUMGrayMedium] (n0) -- (n4);
\draw[dashed,TUMGrayMedium] (n1) -- (n5);
\draw[dashed,TUMGrayMedium] (n5) -- (n2);
\draw[dashed,TUMGrayMedium] (n2) -- (n4);
\coordinate (ni0) at (0.5,0.5,0.5) ;
\coordinate (nb0) at (1.5,0.0,0.0) ;
\coordinate (nb1) at (0.0,1.5,0.0) ;
\coordinate (nb2) at (0.0,0.0,1.5) ;
\coordinate (nB0) at (0.5,0.5,0.0) ;
\coordinate (nB3) at (0.0,0.5,0.5) ;
\coordinate (nB4) at (0.5,0.0,0.5) ;
\coordinate (n0) at (1.0,0.5,1.0) ;
\coordinate (n1) at (1.5,0.0,1.5) ;
\coordinate (n2) at (1.0,0.5,0.5) ;
\coordinate (n3) at (0.5,0.5,1.0) ;
\coordinate (n4) at (0.5,1.0,1.0) ;
\coordinate (n5) at (0.0,1.5,1.5) ;
\coordinate (n6) at (0.5,1.0,0.5) ;
\coordinate (n7) at (1.0,1.0,0.5) ;
\coordinate (n8) at (1.5,1.5,0.0) ;
\coordinate (n9) at (1.5,1.5,1.5) ;
\coordinate (n10) at (0.5,0.0,0.0) ;
\coordinate (n11) at (1.0,0.5,0.0) ;
\coordinate (n12) at (0.5,1.0,0.0) ;
\coordinate (n13) at (0.0,0.5,0.0) ;
\coordinate (n14) at (0.5,0.0,1.0) ;
\coordinate (n15) at (0.0,0.5,1.0) ;
\coordinate (n16) at (0.0,0.0,0.5) ;
\coordinate (n17) at (1.0,0.0,0.5) ;
\coordinate (n18) at (0.0,1.0,0.5) ;
\coordinate (n19) at (0.0,0.0,0.0) ;
\coordinate (n20) at (1.0,0.0,0.0) ;
\coordinate (n21) at (0.0,1.0,0.0) ;
\coordinate (n22) at (0.0,0.0,1.0) ;
\begin{scope}[canvas is plane={O(0.75,0.75,0.5)x(1.75,0.75,0.5)y(0.75,1.75,0.5)}]
	\node[TUMGreen] (fi11a) at (0,0) {};
	\circledarrow{TUMGreen}{fi11a}{0.2}
\end{scope}
\fill[TUMGreen,fill opacity=0.05] (n7.center) -- (n6.center) -- (ni0.center) -- (n2.center) -- cycle;
\begin{scope}[canvas is plane={O(1.29,0.75,0.208)x(1.29,1.75,0.208)y(0.585,0.75,0.915)}]
	\node[TUMGreen] (fi11b) at (0,0) {};
	\circledarrow{TUMGreen}{fi11b}{0.2}
\end{scope}
\fill[TUMGreen,fill opacity=0.05] (n7.center) -- (n2.center) -- (nb0.center) -- (n8.center) -- cycle;
\begin{scope}[canvas is plane={O(0.75,1.29,0.208)x(1.75,1.29,0.208)y(0.75,2.0,-0.499)}]
	\node[TUMGreen] (fi11c) at (0,0) {};
	\circledarrow{TUMGreen}{fi11c}{0.2}
\end{scope}
\fill[TUMGreen,fill opacity=0.05] (n7.center) -- (n8.center) -- (nb1.center) -- (n6.center) -- cycle;
\end{tikzpicture}
    }
    \qquad
    \subfloat[Volume\label{fig:dualVolume}]
    {
        \begin{tikzpicture}[MyPersp,scale=1.3]
\coordinate (nb0) at (0.0,0.0,0.0) ;
\coordinate (nb1) at (1.0,0.0,0.0) ;
\coordinate (nb2) at (0.0,1.0,0.0) ;
\coordinate (nb3) at (1.0,1.0,0.0) ;
\coordinate (nb4) at (0.0,0.0,1.0) ;
\coordinate (nb5) at (1.0,0.0,1.0) ;
\coordinate (nb6) at (0.0,1.0,1.0) ;
\node[circle,fill,inner sep=0pt, minimum size=2mm,TUMOrange] (ni0) at (1.0,1.0,1.0) {};
\coordinate (nB0) at (1.5,1.5,1.5) ;
\coordinate (n0) at (1.5,0.0,0.0) ;
\coordinate (n1) at (0.0,1.5,0.0) ;
\coordinate (n2) at (0.0,0.0,1.5) ;
\coordinate (n3) at (1.5,1.5,0.0) ;
\coordinate (n4) at (1.5,0.0,1.5) ;
\coordinate (n5) at (0.0,1.5,1.5) ;
\draw[TUMGrayMedium] (nb0) -- (nb1);
\draw[TUMGrayMedium] (nb1) -- (nb3);
\draw[TUMGrayMedium] (nb3) -- (nb2);
\draw[TUMGrayMedium] (nb2) -- (nb0);
\draw[TUMGrayMedium] (nb4) -- (nb5);
\draw[TUMGrayMedium] (nb5) -- (ni0);
\draw[TUMGrayMedium] (ni0) -- (nb6);
\draw[TUMGrayMedium] (nb6) -- (nb4);
\draw[TUMGrayMedium] (nb0) -- (nb4);
\draw[TUMGrayMedium] (nb1) -- (nb5);
\draw[TUMGrayMedium] (nb2) -- (nb6);
\draw[TUMGrayMedium] (nb3) -- (ni0);
\draw[TUMGrayMedium] (ni0) -- (nB0);
\draw[TUMGrayMedium] (nB0) -- (n4);
\draw[TUMGrayMedium] (n4) -- (nb5);
\draw[TUMGrayMedium] (nB0) -- (n3);
\draw[TUMGrayMedium] (n3) -- (nb3);
\draw[TUMGrayMedium] (nb6) -- (n5);
\draw[TUMGrayMedium] (n5) -- (nB0);
\draw[dashed,TUMGrayMedium] (nb1) -- (n0);
\draw[dashed,TUMGrayMedium] (n0) -- (n3);
\draw[dashed,TUMGrayMedium] (nb2) -- (n1);
\draw[dashed,TUMGrayMedium] (n1) -- (n3);
\draw[dashed,TUMGrayMedium] (nb4) -- (n2);
\draw[dashed,TUMGrayMedium] (n0) -- (n4);
\draw[dashed,TUMGrayMedium] (n1) -- (n5);
\draw[dashed,TUMGrayMedium] (n5) -- (n2);
\draw[dashed,TUMGrayMedium] (n2) -- (n4);
\coordinate (ni0) at (0.5,0.5,0.5) ;
\coordinate (nb0) at (1.5,0.0,0.0) ;
\coordinate (nb1) at (0.0,1.5,0.0) ;
\coordinate (nb2) at (0.0,0.0,1.5) ;
\coordinate (nB0) at (0.5,0.5,0.0) ;
\coordinate (nB3) at (0.0,0.5,0.5) ;
\coordinate (nB4) at (0.5,0.0,0.5) ;
\coordinate (n0) at (1.0,0.5,1.0) ;
\coordinate (n1) at (1.5,0.0,1.5) ;
\coordinate (n2) at (1.0,0.5,0.5) ;
\coordinate (n3) at (0.5,0.5,1.0) ;
\coordinate (n4) at (0.5,1.0,1.0) ;
\coordinate (n5) at (0.0,1.5,1.5) ;
\coordinate (n6) at (0.5,1.0,0.5) ;
\coordinate (n7) at (1.0,1.0,0.5) ;
\coordinate (n8) at (1.5,1.5,0.0) ;
\coordinate (n9) at (1.5,1.5,1.5) ;
\coordinate (n10) at (0.5,0.0,0.0) ;
\coordinate (n11) at (1.0,0.5,0.0) ;
\coordinate (n12) at (0.5,1.0,0.0) ;
\coordinate (n13) at (0.0,0.5,0.0) ;
\coordinate (n14) at (0.5,0.0,1.0) ;
\coordinate (n15) at (0.0,0.5,1.0) ;
\coordinate (n16) at (0.0,0.0,0.5) ;
\coordinate (n17) at (1.0,0.0,0.5) ;
\coordinate (n18) at (0.0,1.0,0.5) ;
\coordinate (n19) at (0.0,0.0,0.0) ;
\coordinate (n20) at (1.0,0.0,0.0) ;
\coordinate (n21) at (0.0,1.0,0.0) ;
\coordinate (n22) at (0.0,0.0,1.0) ;
\begin{scope}[canvas is plane={O(1.29,0.208,0.75)x(1.29,0.208,1.75)y(0.585,0.915,0.75)}]
	\node[TUMRose] (fi5a) at (0,0) {};
	\circledarrow{TUMRose}{fi5a}{0.2}
\end{scope}
\filldraw[TUMRose,fill opacity=0.05] (n2.center) -- (nb0.center) -- (n1.center) -- (n0.center) -- cycle;
\begin{scope}[canvas is plane={O(0.75,0.5,0.75)x(1.75,0.5,0.75)y(0.75,0.5,1.75)}]
	\node[TUMRose] (fi5b) at (0,0) {};
	\circledarrow{TUMRose}{fi5b}{0.2}
\end{scope}
\filldraw[TUMRose,fill opacity=0.05] (n3.center) -- (ni0.center) -- (n2.center) -- (n0.center) -- cycle;
\begin{scope}[canvas is plane={O(0.75,0.208,1.29)x(1.75,0.208,1.29)y(0.75,-0.499,2.0)}]
	\node[TUMRose] (fi5c) at (0,0) {};
	\circledarrow{TUMRose}{fi5c}{0.2}
\end{scope}
\filldraw[TUMRose,fill opacity=0.05] (n1.center) -- (nb2.center) -- (n3.center) -- (n0.center) -- cycle;
\begin{scope}[canvas is plane={O(0.208,1.29,0.75)x(0.208,1.29,1.75)y(-0.499,2.0,0.75)}]
	\node[TUMRose] (fi6a) at (0,0) {};
	\circledarrow{TUMRose}{fi6a}{0.2}
\end{scope}
\filldraw[TUMRose,fill opacity=0.05] (n4.center) -- (n5.center) -- (nb1.center) -- (n6.center) -- cycle;
\begin{scope}[canvas is plane={O(0.5,0.75,0.75)x(0.5,1.75,0.75)y(0.5,0.75,-0.25)}]
	\node[TUMRose] (fi6b) at (0,0) {};
	\circledarrow{TUMRose}{fi6b}{0.2}
\end{scope}
\filldraw[TUMRose,fill opacity=0.05] (n4.center) -- (n6.center) -- (ni0.center) -- (n3.center) -- cycle;
\begin{scope}[canvas is plane={O(0.208,0.75,1.29)x(0.208,1.75,1.29)y(0.915,0.75,0.585)}]
	\node[TUMRose] (fi6c) at (0,0) {};
	\circledarrow{TUMRose}{fi6c}{0.2}
\end{scope}
\filldraw[TUMRose,fill opacity=0.05] (n4.center) -- (n3.center) -- (nb2.center) -- (n5.center) -- cycle;
\begin{scope}[canvas is plane={O(0.75,0.75,0.5)x(1.75,0.75,0.5)y(0.75,-0.25,0.5)}]
	\node[TUMRose] (fi11a) at (0,0) {};
	\circledarrow{TUMRose}{fi11a}{0.2}
\end{scope}
\filldraw[TUMRose,fill opacity=0.05] (n2.center) -- (ni0.center) -- (n6.center) -- (n7.center) -- cycle;
\begin{scope}[canvas is plane={O(1.29,0.75,0.208)x(1.29,1.75,0.208)y(2.0,0.75,-0.499)}]
	\node[TUMRose] (fi11b) at (0,0) {};
	\circledarrow{TUMRose}{fi11b}{0.2}
\end{scope}
\filldraw[TUMRose,fill opacity=0.05] (n8.center) -- (nb0.center) -- (n2.center) -- (n7.center) -- cycle;
\begin{scope}[canvas is plane={O(0.75,1.29,0.208)x(1.75,1.29,0.208)y(0.75,0.585,0.915)}]
	\node[TUMRose] (fi11c) at (0,0) {};
	\circledarrow{TUMRose}{fi11c}{0.2}
\end{scope}
\filldraw[TUMRose,fill opacity=0.05] (n6.center) -- (nb1.center) -- (n8.center) -- (n7.center) -- cycle;
\begin{scope}[canvas is plane={O(0.75,0.75,1.5)x(1.75,0.75,1.5)y(0.75,1.75,1.5)}]
	\node[TUMRose] (fb0a) at (0,0) {};
	\circledarrow{TUMRose}{fb0a}{0.2}
\end{scope}
\filldraw[TUMRose,fill opacity=0.05] (n9.center) -- (n5.center) -- (nb2.center) -- (n1.center) -- cycle;
\begin{scope}[canvas is plane={O(1.5,0.75,0.75)x(1.5,1.75,0.75)y(1.5,0.75,1.75)}]
	\node[TUMRose] (fb0b) at (0,0) {};
	\circledarrow{TUMRose}{fb0b}{0.2}
\end{scope}
\filldraw[TUMRose,fill opacity=0.05] (n9.center) -- (n1.center) -- (nb0.center) -- (n8.center) -- cycle;
\begin{scope}[canvas is plane={O(0.75,1.5,0.75)x(1.75,1.5,0.75)y(0.75,1.5,-0.25)}]
	\node[TUMRose] (fb0c) at (0,0) {};
	\circledarrow{TUMRose}{fb0c}{0.2}
\end{scope}
\filldraw[TUMRose,fill opacity=0.05] (n9.center) -- (n8.center) -- (nb1.center) -- (n5.center) -- cycle;
\end{tikzpicture}
    }
    \caption{Primal and associated dual cells}
    \label{fig:dualCells}
\end{figure}

The dual 3-complex is defined by construction.
For better visibility, only one dual $j$-cell is drawn in Fig. \ref{fig:dualCells}.
The same procedure is repeated for all other primal $j$-cells.

A barycentric dual is used, as in \cite{Alotto2013}. 
This means, that the dual node is located at the barycentre of the primal volume (Fig. \ref{fig:dualNode}).
Accordingly, a dual edge intersects with its primal face at the barycentre of the face (Fig. \ref{fig:dualEdge})
and the dual face intersects with the primal edge also at the barycentre of the edge (Fig. \ref{fig:dualFace}).
The dual complex is completed with the dual volumes around the primal nodes (Fig. \ref{fig:dualVolume}).

\subsection{Discrete PH Respresentation \label{sec:ph}}
For a structured discrete model of the heat transfer on the foam,
we start with the well-known heat equation with distributed parameters
on a single phase, $\T{x} \in \Omega \subset \mathbb R^3$, $t \in \mathbb R^+_0$,
\begin{align}
	\label{eq:heat-eq}
    c \dot{T}(\T{x},t) &= \lambda \Delta T(\T{x},t).
\end{align}%
\nomenclature[T]{$T$}{Distributed temperature $T(\T{x},t)$}%
\nomenclature[x]{$\T{x}$}{Space variable}%
\nomenclature[t]{$t$}{Time variable}%
\nomenclature[c]{$c$}{Heat capacity}%
\nomenclature[*l]{$\lambda$}{Thermal conductivity}%
$T(\T{x},t)$ denotes the temperature, the heat capacity $c$ and 
the thermal conductivity $\lambda$ are assumed to be constant. 
We rewrite \eqref{eq:heat-eq} in port-Hamiltonian form 
(neglecting for the moment the boundary conditions) 
using the inner energy density $u(\T{x},t)$ 
as state and $T(\T{x},t)$ as co-state/effort%
\footnote{Which is the conjugate quantity w.\,r.\,t. 
the artificial potential $\int_\Omega \frac{1}{c} u^2(\T{x},t) d\T{x}$.}, 
\begin{align}
	\label{eq:heat-eq-structured}
    \begin{bmatrix}
        \dot{u} \\ 
        \T{f} \\
    \end{bmatrix} &=
    \begin{bmatrix}
        0 & -\mathrm{div} \\
        -\mathrm{grad} & \T{0} \\
    \end{bmatrix}
    \begin{bmatrix}
        T \\
        \T{\phi} \\
    \end{bmatrix}.
\end{align}
$\T{\phi}(\T{x},t)$ and $\T{f}(\T{x},t)$ denote the vectors of heat flux
and the temperature gradient as the thermodynamic driving force.
The model is completed with the constitutive laws
\begin{align}
    \T{\phi} &= \lambda \T{f} & u &= cT.
    \label{eq:constEq}
\end{align}

The discrete model is found by integrating the equations
over the appropriate $j$-chains of the dual and the primal complex, respectively,
as indicated in Table \ref{tab:integrationDomains} with superscript s or f referring
to the solid or the fluid phase.

\begin{table}[!ht]
	\label{tab:integrationDomains}
	\centering
    \caption{$j$-chains and associated quantities}    
    \begin{tabular}{ll}
        $j$-chain & (Integral) physical quantity \\
        \hline
        Primal node $n_{k}$     & Temperature $T_{k}^\mathrm{s/f}$ \\
        Primal edge $e_{k}$     & Driving force (temperature difference) $F_{k}^\mathrm{s/f}$ \\
        Dual face $\hat{f}_{k}$ & Heat flow rate $\hat{\Phi}_{k}^\mathrm{s/f}$ \\
        Dual volume $\hat{v}_{k}$ & Energy $\hat{U}_{k}^\mathrm{s/f}$ \\
        \hline
    \end{tabular} 
\end{table}

If $k$ is the index for a dual control volume, and the set $\mathcal I(k)$ contains the indices of the boundary faces, the discrete energy balance on such a control volume can be written for both the solid and the fluid phase as
\begin{subequations}
	\label{eq:discreteEnergyBalance}
\begin{align}
	\frac{\partial}{\partial t}\hat U_k^\mathrm{s} &= -\sum_{l\in \mathcal I(k)} \hat \Phi_{k,l}^\mathrm{s} - \hat \Phi_k^\mathrm{sf}  \\
	\frac{\partial}{\partial t}\hat U_k^\mathrm{f} &= -\sum_{l\in \mathcal I(k)} \hat \Phi_{k,l}^\mathrm{f} + \hat \Phi_k^\mathrm{sf}.
\end{align}\end{subequations}
The heat flow $\hat{\Phi}_{k}^\mathrm{sf}$ represents the heat transfer between both phases. The temperature differences along a strut (index $k$, $1$ and $2$ refer to the start and end node) for both phase, as well as between both phases are
\begin{subequations}
	\label{eq:discreteDrivingForces}
\begin{align}
	F_k^\mathrm{s} &=-(T_{k,2}^\mathrm{s}-T_{k,1}^\mathrm{s}), \quad
	F_k^\mathrm{f} =-(T_{k,2}^\mathrm{f}-T_{k,1}^\mathrm{f})\\
	F_k^\mathrm{sf} &=T_{k}^s-T_{k}^\mathrm{f}.
\end{align}\end{subequations}
Finally, the discrete approximations of the constitutive equations \eqref{eq:constEq} for both phases, together with the heat transfer model between both phases are
\begin{subequations}\begin{align}
	\hat{\Phi}_k^\mathrm{s} &= \frac{\lambda A_k^\mathrm{s} F_k^\mathrm{s} }{|\T{r}_{k,2}-\T{r}_{k,1}|}, \quad 
    \hat{\Phi}_k^\mathrm{f} = \frac{\lambda A_k^\mathrm{f} F_k^\mathrm{f} }{|\T{r}_{k,2}-\T{r}_{k,1}|}\\ 
    \hat{\Phi}_k^\mathrm{sf} &= \alpha A_k^\mathrm{sf} F_k^\mathrm{sf}
\end{align}
\begin{align}
    \hat{U}_k^\mathrm{s} &= V_k^\mathrm{s} c^\mathrm{s} T_k^\mathrm{s},&
    \hat{U}_k^\mathrm{f} &= V_k^\mathrm{f} c^\mathrm{f} T_k^\mathrm{f}.
\end{align}\end{subequations}
The discrete geometry parameters (note that \eqref{eq:discreteEnergyBalance} and \eqref{eq:discreteDrivingForces} contain only topological information) are given in Table \ref{tab:geomParam}.

\begin{table}[!ht]
	\centering
    \caption{Geometry parameters}
    \label{tab:geomParam}
    \begin{tabular}{ll}
        Parameter & Definition \\
        \hline
        $\T{r}_k$ & Position vector of node $n_k$ \\
        $A_k^\mathrm{s/f}$ & Solid / fluid part of the area of  $f_k$ \\
        $A_k^\mathrm{sf}$ & Contact area of the phases in $v_k$ \\
        $V_k^\mathrm{s/f}$ & Solid / fluid part of the volume of $v_k$ \\
        \hline
    \end{tabular} 
\end{table}

To obtain a numerical model of the heat transfer in the complete foam, we collect the whole set of variables $\hat{U}_k$, $F_k$, $T_k$ and $\hat{\Phi}_k$ 
in the vectors $\T{\hat{U}_\mathrm{i/b}}$, $\T{F}_\mathrm{i/b}$, $\T{T}_\mathrm{i/b}$ and $\T{\hat{\Phi}}_\mathrm{i/b}$, which represent inner / border co-chains as algebraically dual objects to the $j$-chains of the primal and the dual complex\footnote{For a given $j-1$-co-chain $c^{j-1}$, which contains the integral values of a quantity over $j-1$-chains, the duality pairing, see \cite{seslija2014explicit},
\begin{align}
    \langle c^{j-1},\partial_j c_j \rangle &= \langle \mathrm{d}^j c^{j-1} , c_j \rangle
\end{align}
defines the co-boundary operator $\mathrm{d}^j$. The sequence of spaces of co-chains and co-boundary operators defines a co-chain complex 
\begin{align}
    \label{eq:coboundary}
    C^0(K,\mathbb{R}) \overset{\mathrm{d}^1}{\longrightarrow} 
    C^1(K,\mathbb{R}) \overset{\mathrm{d}^2}{\longrightarrow} \ldots
    \overset{\mathrm{d}^n}{\longrightarrow} C^n(K,\mathbb{R}). 
\end{align}
}.

The result is the following system of equations, 
where $\hat{\T{\mathrm{d}}}^3_\mathrm{ii/bi} = - (\T{\mathrm{d}}^1_\mathrm{ii/bi})^T$ and $\T{\mathrm{d}}^1_\mathrm{ii/ib}$ 
denote the co-incidence matrices (i.\,e. the transposed boundary matrices) 
between faces and volumes on the dual complex and nodes and edges on the primal complex, 
respectively\footnote{For the relations of co-incidence matrices between the dual 
complexes, see \cite{Kotyczka2017Dis} or \cite{seslija2014explicit}.}.

\begin{multline}
    \begin{bmatrix} 
        \T{\dot{\hat{U}}}^\mathrm{s}_\mathrm{i} \\ 
        \T{\dot{\hat{U}}}^\mathrm{f}_\mathrm{i} \\
        \T{F}^\mathrm{s}_\mathrm{i}\\ 
        \T{F}^\mathrm{f}_\mathrm{i}\\ 
        \T{F}^\mathrm{sf}_\mathrm{i}
    \end{bmatrix} =
    \begin{bmatrix}
        \T{0} & \T{0} &  (-\T{d}^1_\mathrm{ii})^T & \T{0} &  \T{I} \\
        \T{0} & \T{0} & \T{0} & (-\T{d}^1_\mathrm{ii})^T  &  -\T{I} \\
        \T{d}^1_\mathrm{ii} & \T{0} & \T{0} & \T{0}& \T{0}\\
        \T{0} & \T{d}^1_\mathrm{ii} & \T{0}& \T{0}& \T{0}\\
        -\T{I} & \T{I} & \T{0}& \T{0}& \T{0}
    \end{bmatrix}
    \begin{bmatrix} 
        \T{T}^\mathrm{s}_\mathrm{i} \\ 
        \T{T}^\mathrm{f}_\mathrm{i} \\ 
        \T{\hat{\Phi}}^\mathrm{s}_\mathrm{i}\\ 
        \T{\hat{\Phi}}^\mathrm{f}_\mathrm{i}\\ 
        \T{\hat{\Phi}}^\mathrm{sf}_\mathrm{i}
    \end{bmatrix}
    \\
    + \begin{bmatrix}
        \T{0} & \T{0} &  (-\T{d}^1_\mathrm{ib})^T& \T{0}\\
        \T{0} & \T{0} & \T{0} & (-\T{d}^1_\mathrm{ib})^T \\
        \T{d}^1_\mathrm{ib} & \T{0} & \T{0} & \T{0}\\
        \T{0} & \T{d}^1_\mathrm{ib} & \T{0}& \T{0}\\
        \T{0}& \T{0}& \T{0}& \T{0} \\
    \end{bmatrix}
    \begin{bmatrix} 
        \T{T}^\mathrm{s}_\mathrm{b} \\
        \T{T}^\mathrm{f}_\mathrm{b} \\ 
        \T{\hat{\Phi}}^\mathrm{s}_\mathrm{b}\\ 
        \T{\hat{\Phi}}^\mathrm{f}_\mathrm{b} 
    \end{bmatrix}
\end{multline}
The subscripts $i$ and $b$ denote the locations 
(in the interior or at the boundary) of the $j$-chains, 
on which the discrete quantities are defined as presented in the previous subsections. Note that the skew-symmetry of the first matrix mimics the formal skew-adjointness of the matrix operator in \eqref{eq:heat-eq-structured}.

The model is again completed by the constitutive laws
\begin{subequations}\begin{align}
    \T{\hat{U}}^\mathrm{s/f} &= \T{C}^\mathrm{s/f} \T{T}^\mathrm{s/f} \\
    \T{\hat{\Phi}}^\mathrm{s/f} &= \T{\Lambda}^\mathrm{s/f} \T{F}^\mathrm{s/f}
\end{align}\end{subequations}
with the diagonal matrices
\begin{subequations}\begin{align}
    \T{C}^\mathrm{s/f} &= \mathrm{diag}(V_k^\mathrm{s/f} c^\mathrm{s/f}) \\
    \T{\Lambda}^\mathrm{s/f} &= \mathrm{diag}\left(\frac{\lambda A_k^\mathrm{s/f} }{|\T{r}_{k,2}-\T{r}_{k,1}|}\right)
\end{align}\end{subequations}

\section{Implementation}

\begin{figure}[!ht]
\def\dualX{5}
\def\dualXC{3.7}


\def\NodeY{-2}
\def\EdgeY{-3.9}
\def\FaceY{-5.9}
\def\VolumeY{-7.9}
\def\ComplexY{-9.5}

\begin{tikzpicture}

\umlemptyclass[x=0,%
               y=0,%
               text=TUMGrayMedium,%
               draw=TUMGrayMedium]%
               {Cell}


\umlemptyclass[x=\dualXC,y=0,%
               text=TUMGrayMedium,%
               draw=TUMGrayMedium]%
               {DualCell}
\umlinherit{DualCell}{Cell}


\umlemptyclass[x=0.6,%
               y=\NodeY,%
               text=TUMOrange,%
               draw=TUMOrange]%
               {Node}

\umlinherit[anchor1=90,%
            anchor2=258,%
            geometry=|-|,%
            color=TUMOrange]%
            {Node}{Cell}


\umlemptyclass[x=\dualX,%
               y=\NodeY,%
               text=TUMOrange,%
               draw=TUMOrange]%
               {DualNode}

\umlinherit[anchor1=180,%
            anchor2=270,%
            geometry=-|,%
            color=TUMOrange]%
            {DualNode}{DualCell}


\umlemptyclass[x=0.6,%
               y=\EdgeY,%
               text=TUMBlue,%
               draw=TUMBlue]{Edge}

\umlinherit[anchor1=180,%
            anchor2=245,%
            geometry=-|,%
            color=TUMBlue]%
            {Edge}{Cell}

\umlaggreg[mult1=2..*,%
           mult2=2,%
           color=TUMBlue]%
          {Edge}{Node}


\umlemptyclass[x=\dualX,%
               y=\EdgeY,%
               text=TUMBlue,%
               draw=TUMBlue]{DualEdge}

\umlinherit[anchor1=180,%
            anchor2=255,%
            geometry=-|,%
            color=TUMBlue]%
            {DualEdge}{DualCell}

\umlaggreg[mult1=2..*,%
           mult2=2,%
           color=TUMBlue]%
          {DualEdge}{DualNode}


\umlemptyclass[x=0.6,%
               y=\FaceY,%
               text=TUMGreen,%
               draw=TUMGreen]{Face}

\umlinherit[anchor1=180,%
            anchor2=234,%
            geometry=-|,%
            color=TUMGreen]%
            {Face}{Cell}

\umlaggreg[mult1=2..*,%
           mult2=3..*,%
           color=TUMGreen]%
          {Face}{Edge}


\umlemptyclass[x=\dualX,%
               y=\FaceY,%
               text=TUMGreen,%
               draw=TUMGreen]
              {DualFace}

\umlinherit[anchor1=180,%
            anchor2=241,%
            geometry=-|,%
            color=TUMGreen]%
           {DualFace}{DualCell}

\umlaggreg[mult1=2..*,%
           mult2=3..*,%
           color=TUMGreen]%
          {DualFace}{DualEdge}


\umlemptyclass[x=0.6,%
               y=\VolumeY,%
               text=TUMRose,%
               draw=TUMRose]{Volume}

\umlinherit[anchor1=180,%
            anchor2=225,%
            geometry=-|,%
            color=TUMRose]%
           {Volume}{Cell}

\umlaggreg[mult1 = 1..2,%
           mult2 = 4..*,%
           color=TUMRose]%
          {Volume}{Face}


\umlemptyclass[x=\dualX,%
               y=\VolumeY,%
               text=TUMRose,%
               draw=TUMRose]
              {DualVolume}

\umlinherit[anchor1=180,%
            anchor2=230,%
            geometry=-|,%
            color=TUMRose]%
           {DualVolume}{DualCell}

\umlaggreg[mult1 = 1..2,%
           mult2 = 4..*,%
           color=TUMRose]%
           {DualVolume}{DualFace}


\umlemptyclass[x=1.2,%
               y=\ComplexY]%
               {PrimalComplex}

\umlaggreg[anchor1 = 25,%
           anchor2 = east,%
           mult1 = 1,%
           pos1 = 0.05,%
           mult2 = 1..*,%
           pos2 = 1.7,%
           geometry=|-]%
          {PrimalComplex}{Node}

\umlaggreg[anchor1 = 28,%
           anchor2 = east,%
           mult2 = 1..*,%
           pos2 = 1.65,%
           geometry=|-]%
          {PrimalComplex}{Edge}

\umlaggreg[anchor1 = 32,%
           anchor2 = east,%
           mult2 = 1..*,%
           pos2 = 1.6,%
           geometry=|-]%
          {PrimalComplex}{Face}

\umlaggreg[anchor1 = 37,%
           anchor2 = east,%
           mult2 = 1..*,%
           pos2 = 1.5,%
           geometry=|-]%
          {PrimalComplex}{Volume}


\umlemptyclass[x=5.9,%
               y=\ComplexY]%
               {DualComplex}

\umlaggreg[anchor1 = 25,%
           anchor2 = east,%
           mult1 = 1,%
           pos1 = 0.05,%
           mult2 = 1..*,%
           pos2 = 1.7,%
           geometry=|-]%
          {DualComplex}{DualNode}

\umlaggreg[anchor1 = 28,%
           anchor2 = east,%
           mult2 = 1..*,%
           pos2 = 1.65,%
           geometry=|-]%
          {DualComplex}{DualEdge}

\umlaggreg[anchor1 = 32,%
           anchor2 = east,%
           mult2 = 1..*,%
           pos2 = 1.6,%
           geometry=|-]%
          {DualComplex}{DualFace}

\umlaggreg[anchor1 = 37,%
           anchor2 = east,%
           mult2 = 1..*,%
           pos2 = 1.4,%
           geometry=|-]%
          {DualComplex}{DualVolume}

\end{tikzpicture}
\caption{Simplified UML diagram}
\label{fig:uml}
\end{figure}
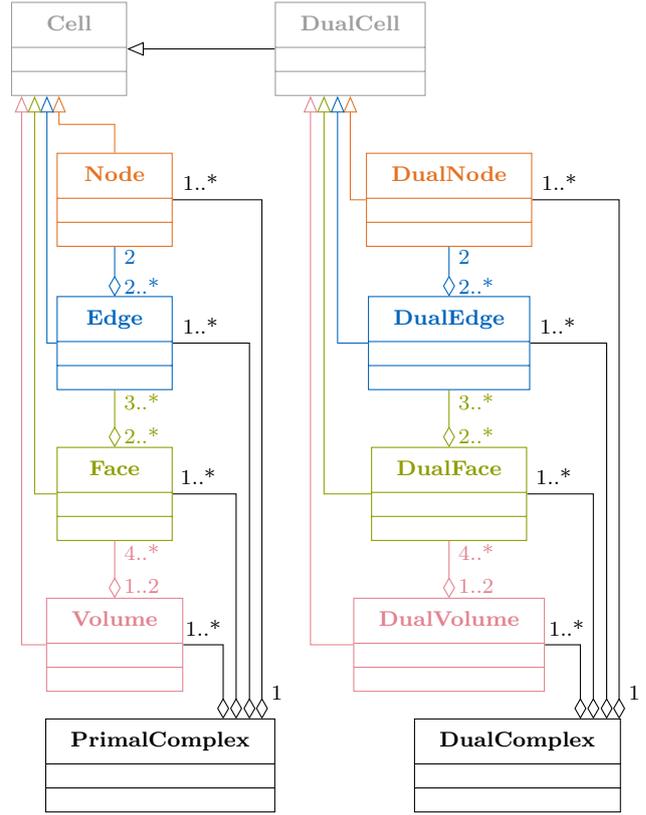

For the implementation of the $3$-complexes and their $j$-cells, 
we chose an objected oriented approach using the programming language \texttt{Python}.
The goal of this implementation is to represent the relations between $j$-cells in the code.
The general structure of the core classes is shown as a UML diagram in Fig. \ref{fig:uml}.
For better treatment of $j$-cells with reverse orientation, 
the implementation includes some more classes than shown,
but they follow the same architecture.

To avoid redundancy, the classes of all $j$-cells inherit from a \texttt{Cell} class
where common properties like numbering or labeling are implemented.
\texttt{Node}, \texttt{Edge}, \texttt{Face} and \texttt{Volumes} classes must be instantiated from top to bottom,
since every class needs an aggregation of its predecessor.
This approach relates to the application of the co-boundary operator as in \eqref{eq:coboundary}.

All objects of $j$-cells are collected in an instance of the \texttt{PrimalComplex} class,
where the classification is implemented and the incidence matrices are calculated.
The \texttt{DualComplex} automatically generates all dual $j$-cells.

\section{Numerical example}
\begin{figure}[!ht]
    \centering
    \includegraphics[width=4cm]{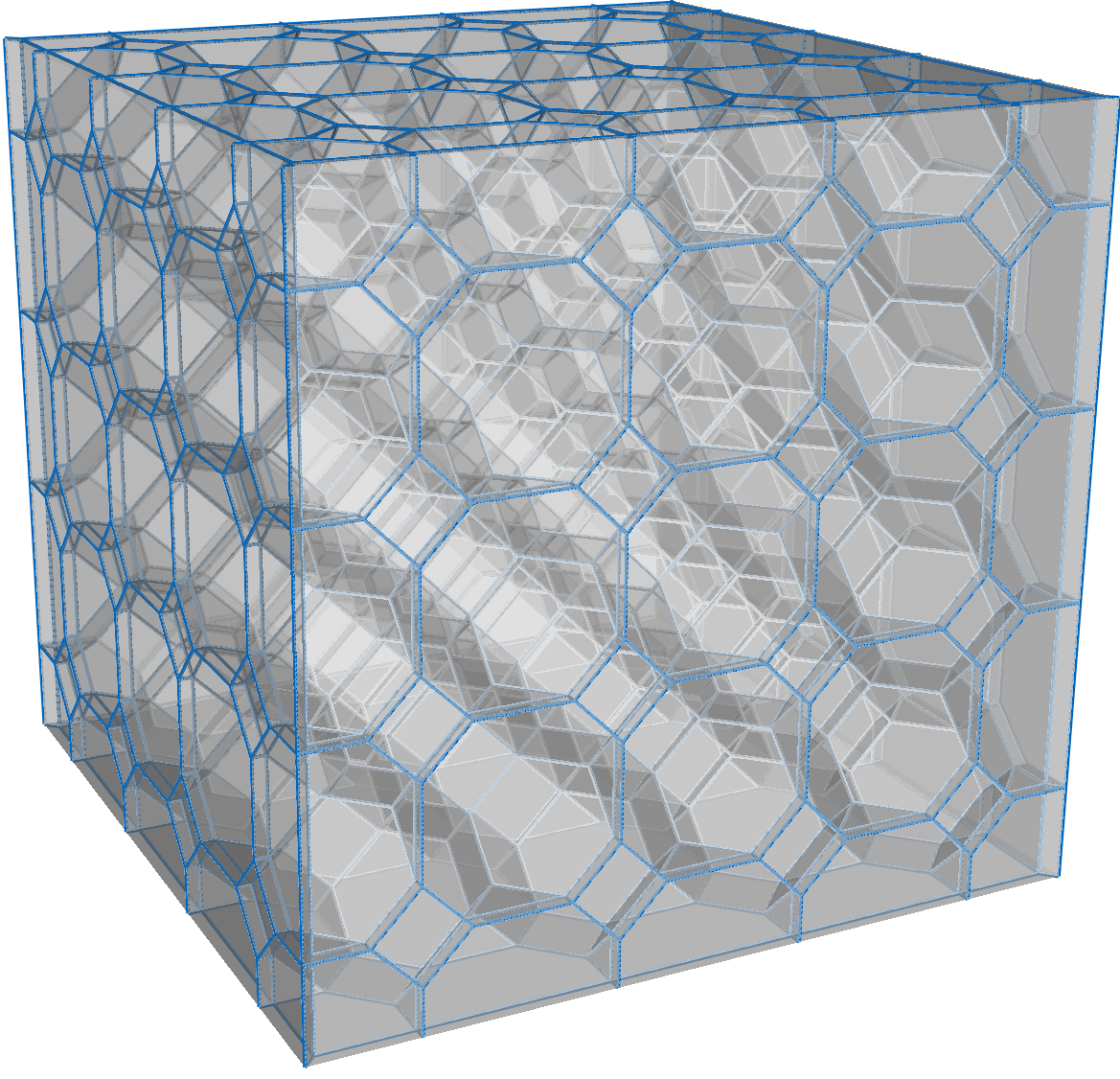}
    \caption{Geometry of the example foam}
    \label{fig:exampleCells}
\end{figure}

\begin{figure}[!ht]
    \hfill
    \subfloat[\acf{dbc}\label{fig:dbc}]
    {
        \def\bc{DBC}
        \ifdefined\bc\else\def\bc{none}\fi

\def\nbcText{NBC}
\def\dbcText{DBC}

\def\myX{2}

\begin{tikzpicture}[MyPersp,scale=1]
\coordinate (nb0) at (0.0,0.0,0.0);
\coordinate (nb1) at (1.0,0.0,0.0);
\coordinate (nb2) at (0.0,1.0,0.0);
\coordinate (nb3) at (1.0,1.0,0.0);
\coordinate (nb4) at (0.0,0.0,1.0);
\coordinate (nb5) at (1.0,0.0,1.0);
\coordinate (nb6) at (0.0,1.0,1.0);
\coordinate (ni0) at (1.0,1.0,1.0);
\coordinate (nm0) at (0.5,0.5,0.0); 
\coordinate (nm1) at (0.5,0.5,1.0); 

\ifx\nbcText\bc
\begin{scope}[canvas is plane={O(1,0.5,0.5)x(0,0.5,0.5)y(1,1.5,0.5)}]
	\blockArrow[TUMGreen]{0,0}{0.5}
\end{scope}
\begin{scope}[canvas is plane={O(0.5,0,0.5)x(0.5,1,0.5)y(1.5,0,0.5)}]
	\blockArrow[TUMGreen]{0,0}{0.5}
\end{scope}
\fi

\draw (nb0) -- (nb1);
\draw (nb1) -- (nb3);
\draw (nb3) -- (nb2);
\draw (nb2) -- (nb0);
\draw (nb4) -- (nb5);
\draw (nb5) -- (ni0);
\draw (ni0) -- (nb6);
\draw (nb6) -- (nb4);
\draw (nb0) -- (nb4);
\draw (nb1) -- (nb5);
\draw (nb2) -- (nb6);
\draw (nb3) -- (ni0);

\ifx\dbcText\bc

\node[circle,fill,inner sep=0pt, minimum size=2mm,TUMOrange] at (nb0) {};
\node[circle,fill,inner sep=0pt, minimum size=2mm,TUMOrange] at (nb1) {};
\node[circle,fill,inner sep=0pt, minimum size=2mm,TUMOrange] at (nb2) {};
\node[circle,fill,inner sep=0pt, minimum size=2mm,TUMOrange] at (nb3) {};
\node[circle,fill,inner sep=0pt, minimum size=2mm,TUMOrange] at (nb4) {};
\node[circle,fill,inner sep=0pt, minimum size=2mm,TUMOrange] at (nb5) {};
\node[circle,fill,inner sep=0pt, minimum size=2mm,TUMOrange] at (nb6) {};
\node[circle,fill,inner sep=0pt, minimum size=2mm,TUMOrange] at (ni0) {};
\node[circle,fill,inner sep=0pt, minimum size=2mm,TUMOrange] at (nm0) {};
\node[circle,fill,inner sep=0pt, minimum size=2mm,TUMOrange] at (nm1) {};

\fill[TUMOrange, fill opacity=0.2] (nb0) -- (nb1) -- (nb3) -- (nb2);
\fill[TUMOrange, fill opacity=0.2] (nb4) -- (nb5) -- (ni0) -- (nb6);
\fi

\ifx\nbcText\bc
\fill[TUMGreen, fill opacity=0.2] (nb0) -- (nb1) -- (nb5) -- (nb4); 
\fill[TUMGreen, fill opacity=0.2] (nb1) -- (nb3) -- (ni0) -- (nb5); 
\fill[TUMGreen, fill opacity=0.2] (nb2) -- (nb3) -- (ni0) -- (nb6); 
\fill[TUMGreen, fill opacity=0.2] (nb2) -- (nb0) -- (nb4) -- (nb6);

\begin{scope}[canvas is plane={O(0,0.5,0.5)x(1,0.5,0.5)y(0,1.5,0.5)}]
	\blockArrow[TUMGreen]{0,0}{0.5}
\end{scope}
\begin{scope}[canvas is plane={O(0.5,1,0.5)x(0.5,0,0.5)y(1.5,1,0.5)}]
	\blockArrow[TUMGreen]{0,0}{0.5}
\end{scope}

\fi

\end{tikzpicture}

    }
    \hfill
    \subfloat[\acf{nbc}\label{fig:nbc}]
    {
        \def\bc{NBC}
        \ifdefined\bc\else\def\bc{none}\fi

\def\nbcText{NBC}
\def\dbcText{DBC}

\def\myX{2}

\begin{tikzpicture}[MyPersp,scale=1]
\coordinate (nb0) at (0.0,0.0,0.0);
\coordinate (nb1) at (1.0,0.0,0.0);
\coordinate (nb2) at (0.0,1.0,0.0);
\coordinate (nb3) at (1.0,1.0,0.0);
\coordinate (nb4) at (0.0,0.0,1.0);
\coordinate (nb5) at (1.0,0.0,1.0);
\coordinate (nb6) at (0.0,1.0,1.0);
\coordinate (ni0) at (1.0,1.0,1.0);
\coordinate (nm0) at (0.5,0.5,0.0); 
\coordinate (nm1) at (0.5,0.5,1.0); 

\ifx\nbcText\bc
\begin{scope}[canvas is plane={O(1,0.5,0.5)x(0,0.5,0.5)y(1,1.5,0.5)}]
	\blockArrow[TUMGreen]{0,0}{0.5}
\end{scope}
\begin{scope}[canvas is plane={O(0.5,0,0.5)x(0.5,1,0.5)y(1.5,0,0.5)}]
	\blockArrow[TUMGreen]{0,0}{0.5}
\end{scope}
\fi

\draw (nb0) -- (nb1);
\draw (nb1) -- (nb3);
\draw (nb3) -- (nb2);
\draw (nb2) -- (nb0);
\draw (nb4) -- (nb5);
\draw (nb5) -- (ni0);
\draw (ni0) -- (nb6);
\draw (nb6) -- (nb4);
\draw (nb0) -- (nb4);
\draw (nb1) -- (nb5);
\draw (nb2) -- (nb6);
\draw (nb3) -- (ni0);

\ifx\dbcText\bc

\node[circle,fill,inner sep=0pt, minimum size=2mm,TUMOrange] at (nb0) {};
\node[circle,fill,inner sep=0pt, minimum size=2mm,TUMOrange] at (nb1) {};
\node[circle,fill,inner sep=0pt, minimum size=2mm,TUMOrange] at (nb2) {};
\node[circle,fill,inner sep=0pt, minimum size=2mm,TUMOrange] at (nb3) {};
\node[circle,fill,inner sep=0pt, minimum size=2mm,TUMOrange] at (nb4) {};
\node[circle,fill,inner sep=0pt, minimum size=2mm,TUMOrange] at (nb5) {};
\node[circle,fill,inner sep=0pt, minimum size=2mm,TUMOrange] at (nb6) {};
\node[circle,fill,inner sep=0pt, minimum size=2mm,TUMOrange] at (ni0) {};
\node[circle,fill,inner sep=0pt, minimum size=2mm,TUMOrange] at (nm0) {};
\node[circle,fill,inner sep=0pt, minimum size=2mm,TUMOrange] at (nm1) {};

\fill[TUMOrange, fill opacity=0.2] (nb0) -- (nb1) -- (nb3) -- (nb2);
\fill[TUMOrange, fill opacity=0.2] (nb4) -- (nb5) -- (ni0) -- (nb6);
\fi

\ifx\nbcText\bc
\fill[TUMGreen, fill opacity=0.2] (nb0) -- (nb1) -- (nb5) -- (nb4); 
\fill[TUMGreen, fill opacity=0.2] (nb1) -- (nb3) -- (ni0) -- (nb5); 
\fill[TUMGreen, fill opacity=0.2] (nb2) -- (nb3) -- (ni0) -- (nb6); 
\fill[TUMGreen, fill opacity=0.2] (nb2) -- (nb0) -- (nb4) -- (nb6);

\begin{scope}[canvas is plane={O(0,0.5,0.5)x(1,0.5,0.5)y(0,1.5,0.5)}]
	\blockArrow[TUMGreen]{0,0}{0.5}
\end{scope}
\begin{scope}[canvas is plane={O(0.5,1,0.5)x(0.5,0,0.5)y(1.5,1,0.5)}]
	\blockArrow[TUMGreen]{0,0}{0.5}
\end{scope}

\fi

\end{tikzpicture}
    }
    \hfill
    \caption{Primal volumes}
    \label{fig:boundaryConditions}
\end{figure}
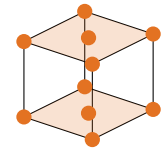
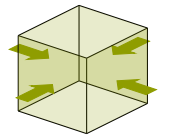

The presented approach is applied to a grid based on Kelvin cells as shown in Fig. \ref{fig:exampleCells}, which has 848 \acp{dof}.
For better replicability, the grid is constructed, so that we can test the numerical method without depending on user settings
in iMorph or the need to compensate possibly occuring defaults in the iMorph result.
On the top and bottom boundary, a \ac{dbc} is applied (Fig. \ref{fig:dbc}).
The other boundaries have a \ac{nbc} (Fig. \ref{fig:nbc}), 
in our case the heat transfer is set to 0, 
meaning it is perfectly isolated at theses boundaries.

The material parameters used in the simulation are given in Table \ref{tab:matParam}.

\renewcommand{\arraystretch}{1.5}
\begin{table}[!ht]
    \caption{Material parameters}
    \label{tab:matParam}
    \begin{tabular}{llrl}
        \hline
        Dimensions & \hspace*{-1cm}$l \times w \times h$ & $\num{40}\times\num{40}\times\num{40}$&\si{\milli \metre}\\
        Mass & $m$ &\num{16.463}  & \si{\gram}\\
        Density of aluminium & $\rho^\mathrm{s}$& \num{2.7e-3}&\si{\gram\per\cubic\milli\meter} \\
        Density of air & $\rho^\mathrm{f}$& \num{1.204e-6}&\si{\gram\per\cubic\milli\meter} \\
        Heat capacity of Al & $c^\mathrm{s}$ &\num{0.897}&\si{\joule\per\gram\per\kelvin} \\
        Heat capacity of air & $c^\mathrm{f}$ &\num{1.005}&\si{\joule\per\gram\per\kelvin} \\
        Thermal conductivity of Al & $\lambda^\mathrm{s}$& \num{0.2}&\si{\watt\per\milli\meter\per\kelvin} \\
        Thermal conductivity of air & $\lambda^\mathrm{f}$& \num{2.6e-5}&\si{\watt\per\milli\meter\per\kelvin} \\
        Heat transfer coefficient & $\alpha$ & \num{1.0e-4}&\si{\watt\per\square\milli\metre\per\kelvin}\\
        Surrogate thermal  diffusivity & $a_\mathrm{eff}$ &\num{1.85}&\si{\square\milli\metre\per\second} \\
        \hline
    \end{tabular} 
\end{table}

\begin{figure}[!ht]
    \centering
    \includegraphics[width=9cm]{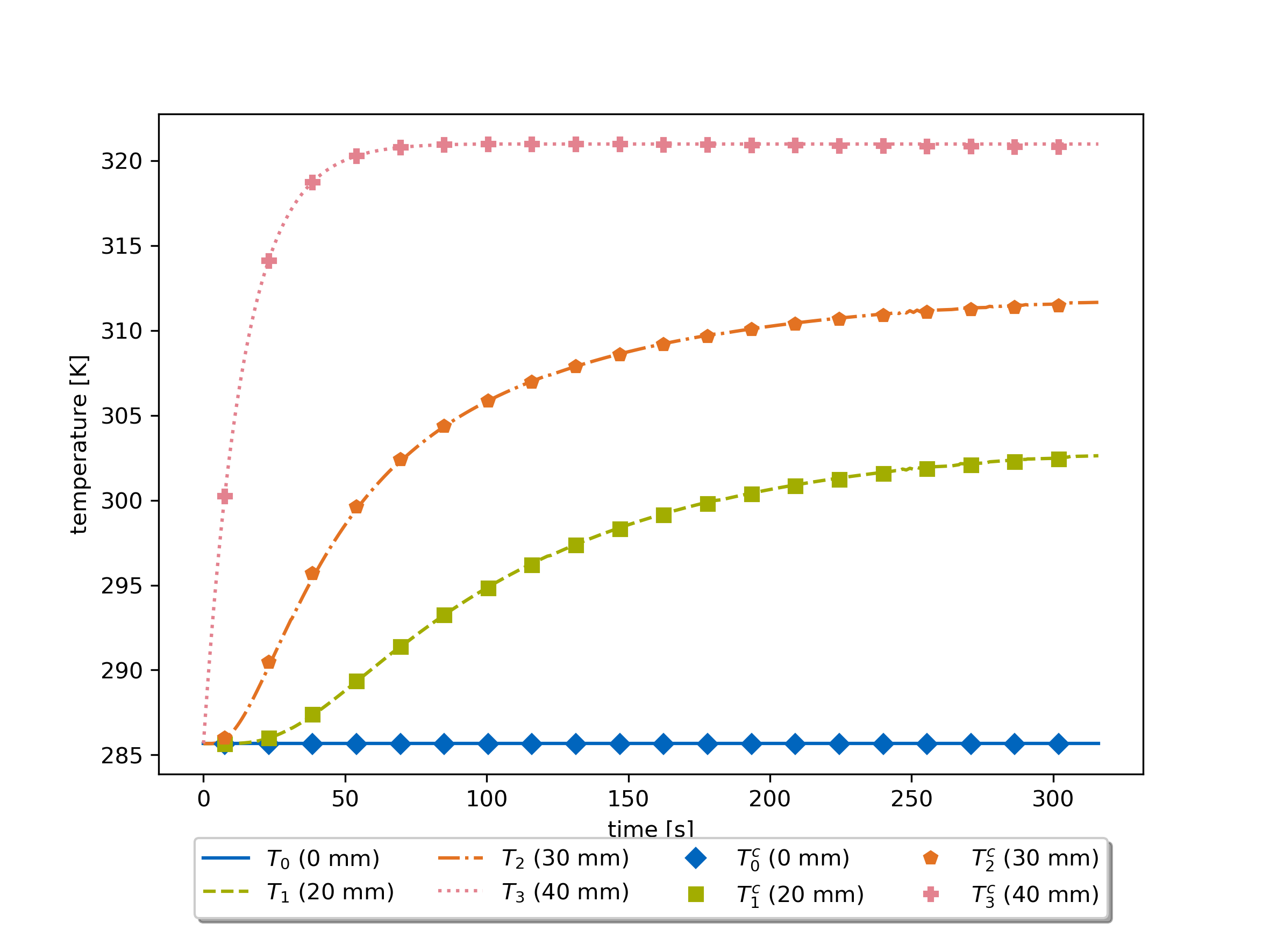}
    \caption{Transient beaviour of the foam}
    \label{fig:examplePlot}
\end{figure}

Fig. \ref{fig:examplePlot} shows the transient behaviour of the temperature on 4 selected nodes.
$T_0$ is the constant temperature at the bottom boundary, while $T_3$ is increased at the top.
$T_1$ and $T_2$ are the temperatures of two nodes at different heights close to the front boundary.

For comparison, 
a \ac{FE} simulation with \num{18081} \acp{dof} was performed with a surrogate parameter for the diffusivity 
$a_{\mathrm{eff}} = \frac{\lambda_\mathrm{eff}}{\rho_{\mathrm{eff}}c_{\mathrm{eff}}}$ 
using FEniCS \citep{AlnaesBlechta2015a}.
The results are shown with markers and the superscript $c$.

\todo[inline]{Write units correctly: time in s, temperature in K}

The perfect matching of both our simulation based on the structured model
with the surrogate \ac{FE} simulation is due to two facts: 
(a) the surrogate diffusivity has been determined by curve fitting and 
(b) did we only consider the ``harmless'' case of pure heat conduction without 
the consideration of convective transport.

\section{Conclusion and Outlook}

We showed a structured approach to obtain a numerical model of heat transfer through metallic open cell foams, in which the separation of topology (expressed in terms of co-incidence matrices) on the one side and geometry and material parameters (constitutive equations) on the other side mimics the \ac{ph} structure of the local PDE model. The model allows to identify macroscopic foam parameters, and can be used for design optimization and (after possible model reduction) for control.

The model structure directly maps to the objects and dependencies of the object oriented python library, which can read topology and geometry data over an interface to the iMorph image processing software. We presented the simulation of a realistic foam model and its comparison to a \ac{FE} simulation with surrogate effective parameters.

At the moment, we work in several directions: (a) the simulation of real foam data and comparison with the experimental data obtained at LGPC Lyon, (b) the integration of convection in the model and (c) improving robustness of our model generation with respect to artefacts like not fully connected graphs from image processing.

\begin{ack}
    The authors cordially thank Jerôme Vicente 
    from University Aix-Marseille for the help with iMorph and in
    particular for implementing modifications in the new iMorph version that allow
    us to directly access all necessary objects and parameters.
\end{ack}

\small
\bibliography{ifacconf} 


\end{document}